\documentclass[aps,twocolumn,prb,amsmath,amssymb,superscriptaddress]{revtex4-1}
\usepackage{url}
\usepackage{graphicx}
\usepackage{color}
\usepackage{natbib}
\usepackage{hyperref}
\usepackage{notes2bib}
\usepackage{mathbbol}
\usepackage{float}
\usepackage{ulem}
\usepackage{tikz}
\usepackage{xcolor}
\usepackage{svg}
\usepackage{siunitx}
\usepackage{dsfont}
\usepackage{ulem}
\usepackage{braket}
\usepackage{relsize}
\usepackage{blkarray}

\begin{document}
\title{Superconductivity in twisted bilayer graphene: possible
pairing symmetries, impurity-induced states and Chern number}
\author{Emile Pangburn}
\affiliation{Institut de Physique Th\'eorique, Universit\'e Paris Saclay, CEA CNRS,
Orme des Merisiers, 91190 Gif-sur-Yvette Cedex, France}
\author{Miguel Alvarado}
\affiliation{Departamento de F\'iƒsica Te\'orica de la Materia Condensada
C-V, Condensed Matter Physics Center (IFIMAC) and Instituto Nicol\'as
Cabrera, Universidad Aut\'onoma de Madrid, E-28049
Madrid, Spain}
\author{Oladunjoye A.~Awoga}
\affiliation{Solid State Physics and NanoLund, Lund University, Box 118, S-221 00 Lund, Sweden}
\author{Catherine P\'epin}
\affiliation{Institut de Physique Th\'eorique, Universit\'e Paris Saclay, CEA CNRS,
Orme des Merisiers, 91190 Gif-sur-Yvette Cedex, France}
\author{Cristina Bena}
\affiliation{Institut de Physique Th\'eorique, Universit\'e Paris Saclay, CEA CNRS,
Orme des Merisiers, 91190 Gif-sur-Yvette Cedex, France}
\date{\today}

\begin{abstract}
We consider the most energetically favorable symmetry-allowed spin-singlet and spin-triplet superconducting pairing symmetries in twisted bilayer graphene at the magic angle, whose normal state physics is described by a six-band effective tight-binding model. We compute the Chern number as a function of the superconducting order parameter strength and the chemical potential and we find a topological phase transition only for the chiral $p+ip'$ superconducting state. Different from the regular graphene systems for which this happens at the van Hove singularity \cite{Adeline2022}, for TBG the topological phase transition arises at the point where the Fermi surface becomes tangent to the boundary of the first Brillouin zone.  For each pairing symmetry we study the formation of subgap impurity states for both scalar and magnetic impurities. We analyze the number of subgap states as well as their spin polarized density of states that we find to exhibit peculiar properties that allows one to distinguish between spin-singlet and triplet pairing. 
Thus only triplet-paired states may exhibit opposite-energy impurity states with the same spin, same as for regular graphene systems\cite{Haurie2022shiba}, moreover we find that this spin may flip at the twist-induced van Hove singularity.
\end{abstract}
\maketitle

\section{Introduction}
Twisted bilayer graphene (TBG) is a two-dimensional material composed of two layers of graphene that are rotated at a small angle relative to each other. When the twist angle is close to a magic value  $\theta\sim 1.05^\circ$, the electronic properties of TBG exhibit spectacular phenomena, such as correlated insulating states, superconductivity, and topological states\cite{cao2018unconventional,po2018origin,lu2019superconductors,balents2020superconductivity,andrei2020graphene,oh2021evidence,cao2021nematicity,christos2020superconductivity,chichinadze2020nematic,wu2020harmonic,yu2021nematicity,fischer2022unconventional}. This has attracted
considerable attention in part because of the exceptional high SC critical temperature, $T_{c}\sim 1 \text{K}$, considering the low electronic density $n\sim 10^{11}\text{cm}^{-2}$, as well as because its strong correlation features that share many similarities with the high-$T_c$ cuprate superconductors. Moreover, because of the extreme tunability of the graphene Moir\'e superlattice, TBG is an excellent candidate to explore various aspects of high-temperature superconductivity.

One of the key questions in the study of TBG is the nature and the pairing symmetry of the emerging superconductivity. Early experiments suggest that the superconductivity in TBG is unconventional, with a pairing symmetry that is inconsistent with the conventional s-wave symmetry\cite{cao2018unconventional}. Subsequent experiments have yielded conflicting results about the nature of this unconventional pairing, and finding the correct pairing in TBG remains a topic of very active research \cite{lake2022pairing,alvarado2023intrinsic,julku2020superfluid,park2021higher,lothman2022nematic}.

A vast range of models for the superconductivity mechanism in TBG have been proposed, relying on either phonon-mediated interactions \cite{lian2019twisted,chou2021Acoustic}, or direct Coulomb interactions between electrons\cite{kennes2018strong}. These theories predict both spin-singlet and spin-triplet order parameters, ranging from s-wave and d-wave to p-wave and f-wave symmetries\cite{BlackSchaffer07, Nandkishore12, Kiesel12,Awoga2017Domain, chichinadze2020nematic, alsharari2022inducing, christos2020superconductivity,alidoust2019symmetry,alidoust2020josephson,thingstad2020phonon,vuvcivcevic2012d,AwogaABC}. This plethora of mechanisms does not provide a definitive conclusion regarding the pairing symmetry in TBG. Thus it is of great importance to design experiments that could distinguish among different pairing symmetries. 

We believe that a partial answer to this question can be given by the study of impurity-induced subgap states \cite{Rusinov1969,Yu1965,Shiba1968}, known as Yu-Rusinov-Shiba for magnetic impurities, which have been widely proposed in many systems such as d-wave high-$T_c$ SCs \cite{hudson2001interplay,zhu2000local,zhang2001marginal,balatsky2006impurity}, iron-based SCs\cite{hirschfeld2015robust,zhou2009impurity,wang2021spin}, p-wave SCs\cite{Kaladzhyan2016CharacterizingUS,guo2017impurity,sau2013bound}, and graphene systems\cite{lothman2014defects,Awoga2018,Haurie2022shiba}, to distinguish among different pairing symmetries. In this work we propose to apply this method also to TBG. The introduction of an impurity can strongly modify the density of states in its vicinity. To investigate this effect, we compute both the unpolarized (DOS) and spin-polarized (SPDOS) local density of states using the T-matrix approach \cite{balatsky2006impurity}. These quantities can in principle be measured by scanning tunneling microscopy (STM). 

Two models, namely, the 2-band model (2B1V) and the 6-band model (6B1V) can be employed to describe TBG \cite{po2019faithful}. The former is straightforward and easy to use, but gives rise to Dirac cones with a net vanishing chirality. However, in TBG, the two Dirac points in the almost flat bands arise from the unperturbed Dirac cones that, despite being in different layers, belong to the same valley, thereby producing a finite net chirality. To address this issue, the 6B1V model includes additional bands while retaining the two distinct flat bands. Therefore, we will concentrate on the 6-band model and only present some results for the 2-band model in the Appendix.

The 6B1V model is a multi-orbital model, so the results depend heavily on which orbitals the impurity affects. While this added complexity makes it challenging to analyze the impact of the impurity, it can also help distinguish between different pairing symmetries that interact differently with the orbitals of the lattice depending on their behavior under point group transformations. Additionally, the spin of the impurity states can provide important insights into the nature (singlet or triplet) of the underlying superconducting order parameter. For example, only triplet-paired states may exhibit opposite-energy impurity states with the same spin, while the impurity states in singlet-paired states always have opposite spins. This phenomenon was also observed in earlier work on graphene \cite{Haurie2022shiba}, and we demonstrate it to be a generic feature of graphene systems with spin-triplet order parameters. Furthermore, the impurity states in triplet-paired SCs may flip spin at the chemical potential corresponding to the twist-induced van Hove singularity, for which the topology of the Fermi surface is modified, and may thus serve as a test for the spin-triplet nature of the SC order parameter, as well as provide insight into the Fermi surface structure.

We have also computed the Chern number for the chiral states $d+id'$ and $p+ip'$ and find that it is independent of any system parameters for the former, while for the latter it changes when the Fermi surface becomes tangent to the boundary of the first Brillouin zone. This transition point is not associated with any visible signature in the density of states, and thus TBG differs from regular graphene systems\cite{Adeline2022} for which the topological phase transition occurs at the Lifshitz transition associated with a van Hove singularity in the density of states.

The structure of the paper is the following. Firstly, in Section \ref{sec:Normal} we detail the tight-binding model used to effectively describe non-interacting TBG at the magic angle. We also review the fermiology and discuss the evolution of the Fermi surface with chemical potential. In Section \ref{sec:SC} we present the model for various $s$-, $p$- and $d$-wave pairing symmetries, and we also detail the method used to compute the DOS, the SPDOS and the Chern number.  The results obtained for the Chern number are presented in Section \ref{sec:Chern}, and the formation of bound states with scalar and magnetic impurities is discussed  in Section \ref{sec:Imp}. Finally, in Section \ref{sec:conclusion}, we summarize our results. In the Appendix, we compare our Chern number results to those obtained using the 2B1V-model, and we provide a detailed explanation of the influence of the pairing symmetries on the number and spin of the subgap states.
%
\section{Normal state of TBG}\label{sec:Normal}
\subsection{Model\label{sec:NormalModel}}
\begin{figure}[H]
\begin{center}
\includegraphics[width=4cm]{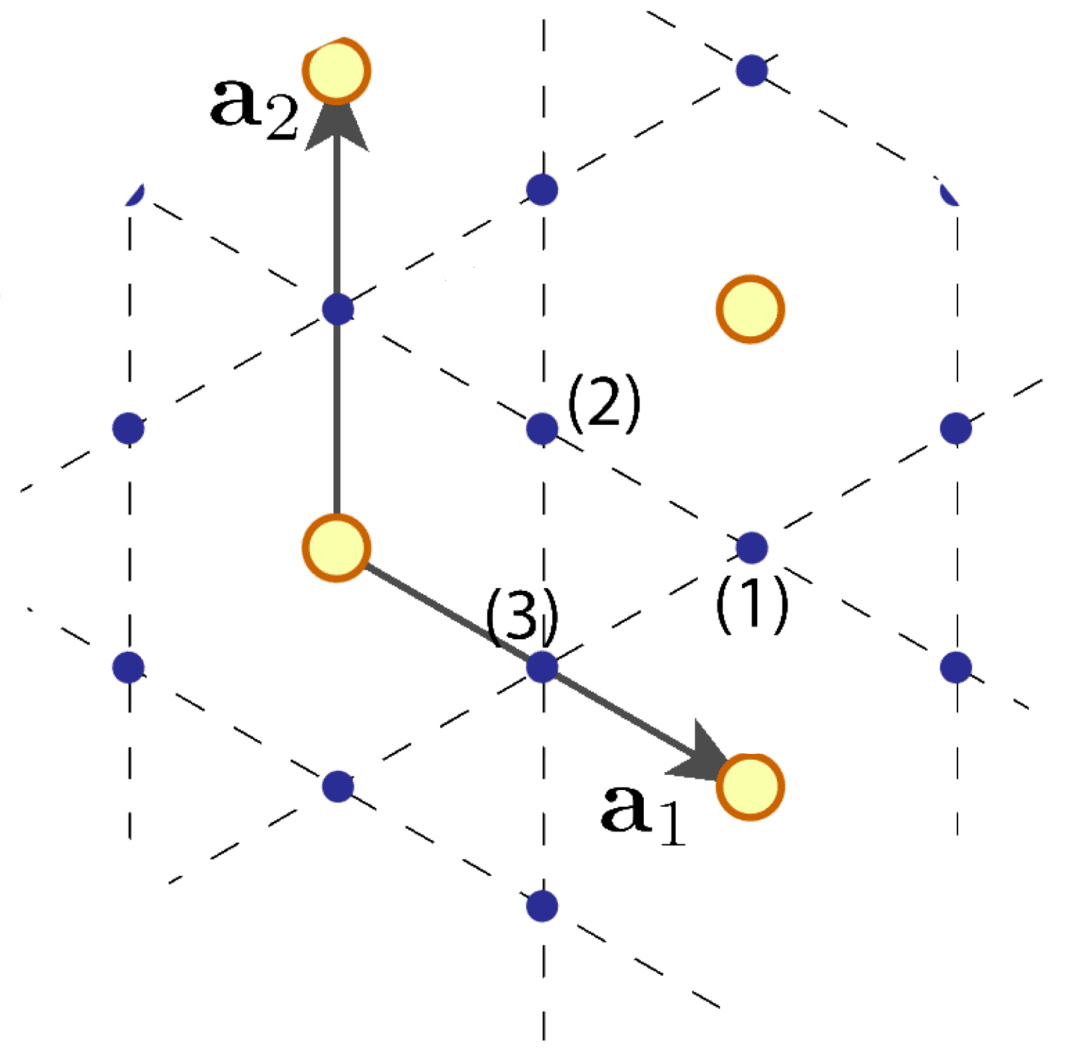} \caption{Figure adapted from Ref. \onlinecite{po2019faithful}. Lattice and conventions.
Blue dots represent the $(\kappa,s)$ orbitals. Yellow dots represent
the $(p_{z},p_{\pm})$ orbitals. We took $a_{1}$ and $a_{2}$ as
basis vectors of the Bravais lattice}
\label{fig:Lattice_6B1V} 
\end{center}
\end{figure}

The physics of TBG at the magic angle is governed by the apparition of two nearly flat bands (per valley and spin). Naively, the minimal model to describe the low energy physics of the system should be a simple two-band model. However, this model cannot capture the non-trivial topology of the TBG normal state \cite{ahn2019failure}, in particular the net finite chirality of the two Dirac points. To avoid this so-called topological obstruction, additional trivial bands need to be added to the flat ones, ensuring that the corresponding orbitals satisfy the relevant symmetries of TBG \cite{bernevig2021twisted}. As a result, we consider an effective 6-band tight-binding model \cite{po2019faithful} that captures the low energy physics of the system of a single valley. In the single valley model of TBG, all inter-valley couplings are neglected, meaning that any symmetry that relates the two valleys of TBG is not a symmetry of this model. Therefore, the only remaining symmetries are $C_{3z}$, $C_{2x}$, and $C_{2z}\mathcal{T}$, which generate the magnetic space group $P6'2'2$\cite{song2019all}. This 6-band model consists of a $p_{z}$ orbital and a pair of $p_{\pm}$ orbitals localized at sites forming a triangular lattice ($\tau$), as well as three $s$ orbitals arranged in a kagome lattice ($\kappa$) as shown in Fig.\ref{fig:Lattice_6B1V}.

We follow Ref.~\cite{po2019faithful} and we use the notation $(p_{z}$, $\tau)$ and ($p_{z}$, $\kappa$) to indicate an orbital located
on the $\tau$, and respectively on the $\kappa$ lattice. We define the fermion operator
that fixes the basis of the Bloch Hamiltonian as 
\begin{equation}
\psi_{\mathbf{k}\sigma}=(\tau_{p_{z}\sigma},\tau_{p_{+}\sigma},\tau_{p_{-}\sigma},\kappa_{s\sigma}^{(1)},\kappa_{s\sigma}^{(2)},\kappa_{s\sigma}^{(3)})^{\top}.\label{eq:Basis_6B1V}
\end{equation}
The form of the non-interacting Bloch Hamiltonian $H_{0}(\mathbf{k})$ in this basis
is given by 
\begin{equation}
\hat{H}_{0}(\mathbf{k})=\begin{pmatrix}H_{p_{z}}+\mu_{p_{z}} & \hat{C}_{p_{\pm}p_{z}}^{\dagger} & \hat{0}\\
\hat{C}_{p_{\pm}p_{z}} & \hat{H}_{p_{\pm}}+\hat{\mu}_{p_{\pm}} & \hat{C}_{\kappa p_{\pm}}^{\dagger}\\
\hat{0} & \hat{C}_{\kappa p_{\pm}} & \hat{H}_{\kappa}+\hat{\mu}_{\kappa},
\end{pmatrix}\label{Eq:H0_6B1Vm}
\end{equation}
and the detailed form of the intra-orbital and inter-orbital coupling terms is given in Appendix \ref{6B1Vmodel}. In what follows, all energies are in units of meV, and all distances are given in units of the lattice constant. To differentiate between scalar and matrix structures in our notation, we employed a hat symbol. For instance, $H_{p_z}$ represents a scalar quantity acting on the $p_z$ orbitals, while $\hat{H}_{p_\pm}$ denotes a 2-by-2 matrix that operates on the $(p_+, p_-)$ orbitals. Note that all the energy values that we use are determined by the choice of parameters in the tight-binding Hamiltonian listed in Table~\ref{tab:parameter}, following Ref.~\onlinecite{alvarado2023intrinsic}.

\subsection{Fermi Surface Topology}\label{sec:NormalFermiology}
In graphene systems, including TBG, it is possible experimentally to shift the Fermi level by gating~\cite{Oostinga2008,Zhang2010Band,Guinea2018Electrostatic}. Here, we briefly review the evolution of the Fermi surface with the chemical potential $\mu$. Understanding the topology of the Fermi surface is essential to understand the Chern number evolution and the spin polarization of the subgap bound states presented in Sections~\ref{sec:Chern} and~\ref{sec:Imp}, respectively.

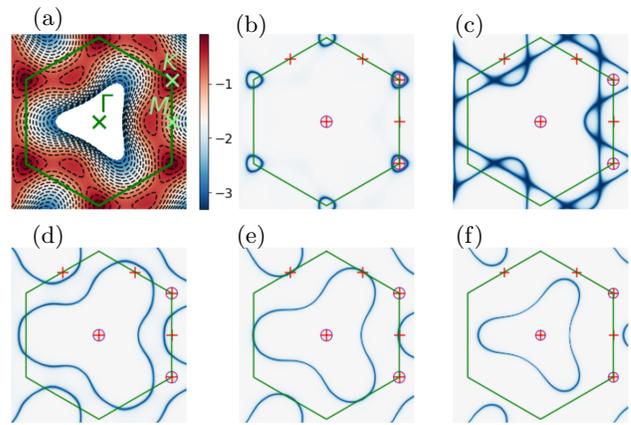
\begin{figure}[!t]
	\begin{center}
		\begin{tikzpicture}
			\node at (0,0) {
				\includegraphics[width=8.5cm]{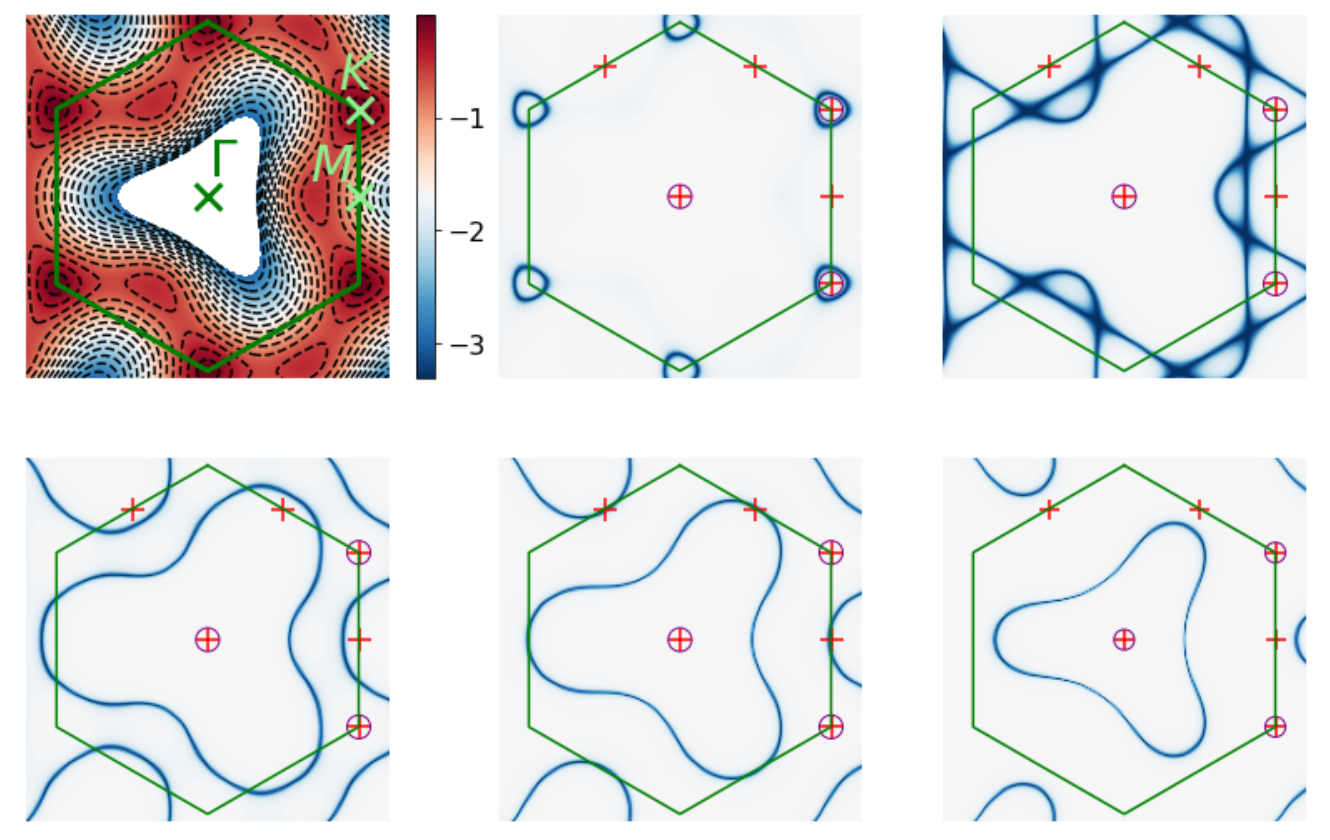}};
			\node at (-3.6,2.8) {(a)}; 
			\node at (-0.9,2.7) {(b)};
			\node at (2.,2.7) {(c)};
			\node at (-3.6,-0.1) {(d)};
			\node at (-0.9,-0.1) {(e)};
			\node at (2,-0.1) {(f)};
		\end{tikzpicture}
	\end{center}
	\caption{Normal state features for the 6B1V model. The green hexagon denotes the first Brillouin zone. In (a) we show the lowest-negative-energy band, the dashed lines denoting isoenergetic contours, and the high-symmetry points $(\Gamma,K,M)$ are indicated in green. (b-f) Fermi-surface contours for $\mu=-0.3$ (b), $\mu=-0.63$ (c), $\mu=-0.9$ (d), $\mu=-1.3$ (e), and $\mu=-2.0$ (f). We denote the special points of the $p+ip'$ pairing by red crosses and the special points of the $d+id'$ pairing by purple X's.}
	\label{fig:FS_6B1V} 
\end{figure}

\begin{figure}[!t]
	\includegraphics[width=6cm]{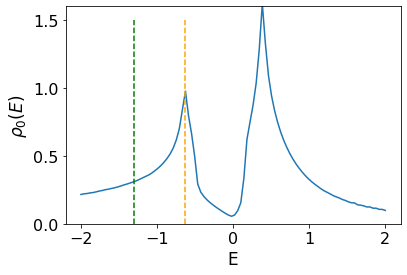} 
	\caption{Unperturbed density of  state of the 6B1V normal state. The orange dashed line indicates the chemical potential $\mu_L\approx -0.63$ corresponding to the Lifshitz transition shown in Fig.~\ref{fig:FS_6B1V} (c). The green dashed line indicates the chemical potential $\mu_T\approx -1.3$ corresponding to the point at which the FS becomes tangent to the BZ boundary, shown in Fig.~\ref{fig:FS_6B1V} (e). Note that a VHS is associated with $\mu_L\approx -0.63$ while no visible DOS feature is associated with $\mu_T\approx -1.3$.}
	\label{fig:NormalStateDOS} 
\end{figure}

We focus on the two bands closest to zero energy since they describe the topology and energetics of the entire system. These bands correspond mainly to the $p_\pm$-orbitals, except close to the $\Gamma$-point where the $p_z$- ($s$-) orbitals have a significant contribution to the negative (positive) band~\cite{po2019faithful}.  In Fig.~\ref{fig:FS_6B1V}(a) we plot the spectrum for the lowest negative energy band. The Brillouin zone (BZ) is denoted by the green hexagon. Note that at low energy the equal-energy contours consist in circles centered around the $K$ points while at higher energy in contours centered around the $\Gamma$ point. Note also the constant-energy region (denoted in red) marking the transition between these two different behaviors. This almost flat region corresponds to a van Hove singularity (VHS) giving rise to a logarithmic divergence in the density of states at $E=-0.63$, see the dashed orange line in Fig.~\ref{fig:NormalStateDOS}.
The VHS singularity can be accessed by doping, i.e modifying the chemical potential so that it aligns with the VHS energy, $E=-0.63$. This is also known as a Lifshitz transition, corresponding to the transition between a regime of low negative doping, for which the Fermi surface consists of two small disconnected pockets centered around the $K$ and $K^\prime$-points (see Fig.~\ref{fig:FS_6B1V}(b)), and a single pocket centered around the $\Gamma$-point, see Fig.~\ref{fig:FS_6B1V}(d-f). At the transition point at $\mu_L=-0.63$, Fig.~\ref{fig:FS_6B1V}(c), it has been noted also a phenomenon of electron-hole conversion\cite{Kim2016Charge,cao2016Superlattice}, and we also observe a spin-inversion of the subgap states, as we will show in what follows. 

Further increasing $\mu$, one reaches the value $\mu_T=-1.3$, for which the Fermi surface becomes tangent to the boundary of the first Brillouin zone, see the dashed green line in Fig.~\ref{fig:FS_6B1V}(e). Interestingly enough, as we will show in what follows, this point corresponds to a topological phase transition and a modification of the Chern number in the $p+ip'$ phase. Note that the transition associated with the FS touching the BZ boundary, indicated by the green line in Fig.~\ref{fig:NormalStateDOS}, does not give rise to any observable signatures in the DOS, different from the Lifshitz transition at $\mu=-0.63$ that is associated with a VHS in the spectrum. 

For an intermediate chemical potential corresponding to the regime between the VHS at $\mu_L=-0.63$ and $\mu_T=-1.3$, having two different Fermi surfaces intersecting the BZ (see Fig.~\ref{fig:FS_6B1V}(d)) has been theorized to enhance unconventional superconducting and density wave instabilities~\cite{isobe2018Unconventional}.  
For very large $\mu$ above $\mu_T=-1.3$, the Fermi surface shrinks more and more around the $\Gamma$-point, see Fig.~\ref{fig:FS_6B1V}(f), without showing any other significant qualitative changes.

Interestingly enough, regular graphene systems show the same Fermi surface evolution with the chemical potential, see e.g. Ref.~\onlinecite{Adeline2022}, i.e a Lifshitz transition and a corresponding VHS, however there is no extra transition since the FS happens to cross the BZ boundary at exactly the VHS. Thus regular graphene has no intermediate $\mu$ regime, which is peculiar to TBG.

\section{Superconducting model}\label{sec:SC}
To model the superconductivity in TBG, one can start with an interacting channel and apply a Hubbard-Stratonovitch transformation \cite{stratonovich1957method}. The resulting saddle-point self-consistent equations can be solved to compute the superconducting order parameter as a function of the interacting potential. However, since there is currently no consensus on the channel responsible for pairing in TBG, we will assume a fixed value $\Delta_0$ for the pairing amplitude in the following analysis. We will explore the different symmetry-allowed pairing symmetries without computing their associated critical temperature and free energy density which could provide information on the favored pairing symmetry.


The unconventional superconductivity observed in TBG is believed to originate from the nearly flat bands that are strongly overlapping with the $(\tau,p_{\pm})$ orbitals at the $K$ points \cite{rademaker2018charge}. As these bands cross the Fermi surface, we introduce a superconducting pairing only on these two orbitals and consider only inter-valley pairing, which is favored by phonon coupling \cite{lian2019twisted} and produces Cooper pairs with zero net momentum. Additionally, we assume that the pairing is orbitally trivial, meaning that if we define $\Delta_{m,m'}(\mathbf{k})\equiv \langle m\mathbf{k}|\hat{\Delta}|m'\mathbf{k}\rangle$, then $\Delta_{m,m'}=\delta_{m,m'}\Delta(\mathbf{k})$, where $\Delta(\mathbf{k})$ belongs to an irreducible representation of the point symmetry group of the triangular lattice, denoted as $D_6$. The classification of these irreducible representations has been studied in the literature, as discussed in \cite{elcoro2017double}.
\begin{table}
\renewcommand*{\arraystretch}{2.2} %
\begin{tabular}{|c|l|l|}
\hline 
$\quad\eta\quad$  & Symmetry  & Form factor $h_{\text{NN}}^{\eta}(\mathbf{k})$ \tabularnewline
\hline 
\;singlet  & \;$s_{\text{ext}}$  & $\frac{2|\Delta_0^{s_{\text{ext}}}|}{\sqrt{6}}\left(2\cos(\frac{3k_{x}}{2})\cos(\frac{k_{y}}{2})+\cos(k_{y})\right)$ \tabularnewline
\;singlet  & \;$d_{x^{2}-y^{2}}$  & $\frac{2|\Delta_0^{d_{x^{2}-y^{2}}}|}{\sqrt{3}}\left(-\cos(\frac{3k_{x}}{2})\cos(\frac{k_{y}}{2})+\cos(k_{y})\right)$ \tabularnewline
\;singlet  & \;$d_{xy}$  & $2|\Delta_0^{d_{xy}}|\sin(\frac{\sqrt{3}}{2}k_{x})\sin(\frac{k_{y}}{2})$ \tabularnewline
\hline 
\; triplet  & \;$p_{y}$  & $-2i|\Delta_0^{p_{y}}|\sin(\frac{\sqrt{3}}{2}k_{x})\cos(\frac{k_{y}}{2})$ \tabularnewline
\; triplet  & \; $p_{x}$  & $-\frac{2i|\Delta_0^{p_{x}}|}{\sqrt{3}}\left(\cos(\frac{3k_{x}}{2})\sin(\frac{k_{y}}{2})+\sin(k_{y})\right)$ \tabularnewline
\hline 
\end{tabular}\caption{Expressions for the form factors corresponding to the various singlet and triplet
symmetries (the distance between two atoms of the $\tau$ lattice
has been set to 1). In our convention for singlet states $\eta=0$, and for triplet states
$\eta\in\{x,y,z\}$, where the index ${x,y,z}$ indicates the spin direction of the
Cooper pair.}
\label{table1} 
\end{table}
In what follows we define $\eta=\{0,x,y,z\}=\{0,\nu\}$. The corresponding SC Hamiltonian can be written as $\hat{H} =\Psi_{\mathbf{k}}^\dagger \hat{H}_{\rm BdG}\Psi_{\mathbf{k}}$, where $\Psi_{\mathbf{k}} = (\psi_{\mathbf{k}\uparrow},\psi_{\mathbf{k}\downarrow},(\psi_{-\mathbf{k}\uparrow})^{\dagger},(\psi_{-\mathbf{k}\downarrow})^{\dagger})^{\top}$ is the $\Psi_{\mathbf{k}}$, and
%
%
the BdG Hamiltonian is given by
\begin{equation}\label{eq:BdG}
	\hat{H}_{\rm BdG} = \begin{pmatrix}\hat{\mathcal{H}}_{0}^K\left(\mathbf{k}\right) & \hat{\Delta}\left(\mathbf{k}\right)\\
	\left(\hat{\Delta}\left(\mathbf{k}\right)\right)^\dagger & -\hat{\mathcal{H}}_{0}^{K'}\left(-\mathbf{k}\right)^*
	\end{pmatrix},\quad
\end{equation}
$\hat{\mathcal{H}}_{0}^{K/K'}=\sigma_0\otimes \hat{H}_{0}^{K/K'}\left(\mathbf{k}\right)$ are the normal state Hamiltonian for respectively the $K/K'$ valley of TBG, $\hat{\Delta}\left(\mathbf{k}\right) = i\sigma_y\otimes \hat{h}_{\Delta}^{0}({\mathbf{k}})$ for spin-singlet pairing $(\eta=0)$  and $\hat{\Delta}\left(\mathbf{k}\right) = \sigma_z\otimes \hat{h}_{\Delta}^{x}({\mathbf{k}})$ for $\eta=x$ spin-triplet pairing. Here $\sigma_\nu$ is the $\nu$-Pauli matrix in spin space. Because the $K$ and $K'$ valleys are related by time-reversal symmetry, we have $\mathcal{H}_0^{K'}(-\mathbf{k})^*=\mathcal{H}_0^{K}(\mathbf{k})$ so finally the BdG Hamiltonian can be written as 
\begin{equation}\label{eq:BdG}
	\hat{H}_{\rm BdG} = \begin{pmatrix}\hat{H}_{0}\left(\mathbf{k}\right) & \hat{\Delta}\left(\mathbf{k}\right)\\
	\left(\hat{\Delta}\left(\mathbf{k}\right)\right)^\dagger & -\hat{H}_{0}\left(\mathbf{k}\right)
	\end{pmatrix},\quad
\end{equation}
where $\hat{H}_{0}(\mathbf{k})$ is the normal state Hamiltonian given in Eq.~(\ref{Eq:H0_6B1Vm}). The
non vanishing entries of the form factors $\hat{h}_{\Delta}^{0}(\mathbf{k})$
and $\hat{h}_{\Delta}^{x}(\mathbf{k})$ are, respectively, $[\hat{h}_{\Delta}^{0}{(\mathbf{k}})]_{2,2}=[\hat{h}_{\Delta}^{0}{(\mathbf{k}})]_{3,3}=\frac{1}{2}h_{NN}^{0}(-\mathbf{k})$ and $[\hat{h}_{\Delta}^{x}{(\mathbf{k}})]_{2,2}=[\hat{h}_{\Delta}^{x}{(\mathbf{k}})]_{3,3}=\frac{1}{2}h_{NN}^{x}(-\mathbf{k})
$.

\subsection{Chern number}
We can compute the Chern number associated with the gapped pairing symmetries $p+ip'$ and $d+id'$. We use the well-known TKNN formula \cite{PhysRevLett.49.405}
\begin{eqnarray}
\mathcal{C} & = & \frac{i}{8\pi^{2}}\int\text{d}{\bf k}\,\text{d}E\;\text{Tr}\bigg[\mathbf{\hat{G}}^{2}({\bf k},E)(\partial_{k_{y}}\hat{H}_{{\bf k}})\mathbf{\hat{G}}({\bf k},E)(\partial_{k_{x}}\hat{H}_{{\bf k}})\nonumber \\
 &  & -\mathbf{\hat{G}}^{2}({\bf k},E)(\partial_{k_{x}}\hat{H}_{{\bf k}})\mathbf{\hat{G}}({\bf k},E)(\partial_{k_{y}}\hat{H}_{{\bf k}})\bigg]~,
\label{eq:TKNN}
\end{eqnarray}

\subsection{Impurity bound states}

We study the consequences of the introduction of a classical point-like (scalar
or magnetic) impurity located on one of the triangular lattice sites. To compute the corresponding variation of the average unpolarised DOS and spin-polarised DOS we use the T-matrix approximation\cite{balatsky2006impurity}.
Using the usual notation for the retarded Green's function of the unperturbed system $\hat{\mathbf{G}}(E,\mathbf{k})=\left[(\hat{E}+i\hat{\delta})-\hat{H}_{\text{BdG}}(\mathbf{k})\right]^{-1}$  the T-matrix is given by
\begin{equation}
\hat{T}(E)=\left[\mathds{1}_{24}-\mathds{V}\cdot\int_{BZ}\frac{d^{2}\mathbf{k}}{A_{\rm BZ}^{}}\mathbf{\hat{G}}(E,\mathbf{k})\right]^{-1}\cdot \mathds{V},\label{eq:Tmatrix}
\end{equation}
where $\delta$ is the quasiparticle-lifetime, that we take to be $\delta=0.01$ in all our calculations without computing it self-consistently. $A_{\rm BZ}^{}$ is the area of the reduced Brillouin zone.

The matrix $\mathds{V}$ corresponds to a point-like (scalar or magnetic) impurity. The parameters $U$ and $\vec{J}=(J_x,J_y,J_z)$ are respectively, the strength of the scalar and magnetic impurity.
We introduce one impurity on a site of the triangular lattice as shown in Fig.~\ref{fig:Lattice_6B1V}. We take the impurity potential to be diagonal in orbital space $(p_z,p_+,p_-)$ so $V$ can be written as,
\begin{equation}
\mathds{V}=\tau^z \otimes[V_{\tau}\otimes[U\sigma_0+\vec{J}.\vec{\sigma}]]
\end{equation}
with $\tau^z$ the Pauli matrices for the particle/hole subspace. The action of the impurity on the orbitals localized on the triangular lattice is encoded in the matrix $V_\tau$, which is a diagonal matrix with only $0$ and $1$ elements. We do not consider any impurities acting on the Kagome lattice as the orbital weight of the Kagome orbitals is everywhere vanishing on the Fermi Surface, and their hybridization with the orbitals on the triangular lattice is small.
The diagonal value of $V_\tau$ is $1$ if the impurity acts on the corresponding orbital and $0$ otherwise.

The physical observables, such as the DOS or spin-polarized DOS near an impurity can be expressed directly in terms of the T-matrix given in Eq.~(\ref{eq:Tmatrix}). This is possible by assuming the dilute limit, where the impurities are well separated from each other. Here, we use $b$ as the orbital index and $\alpha,\beta\in{\uparrow,\downarrow}$ as the spin index. The impurity contribution is given by,
\begin{equation}
\delta S_{\eta}\left(\mathbf{p},E\right)=\frac{-1}{2\pi i}\int_{BZ}\frac{d^{2}\mathbf{q}}{A_{\rm BZ}^{}}\sum\limits _{b,\alpha \beta} g_{b,\alpha\beta}^\eta \left(E,\mathbf{q},\mathbf{p}\right)  \tilde{\sigma}_{\alpha\beta}^\eta,
\label{eq:spin_components}
\end{equation}
where 
\begin{equation}
	\begin{split}
		  g^\eta\left(E,\mathbf{q},\mathbf{p}\right) & =  \mathbf{\hat{G}}\left(E,\mathbf{q}\right)\hat{T}\left(E\right)\mathbf{\hat{G}}\left(E,\mathbf{q+p}\right) \\
		& +s^\eta \mathbf{\hat{G}}^{\star}\left(E,\mathbf{p+q}\right)\hat{T}^{\star}\left(E\right)\mathbf{\hat{G}}^{\star}\left(E,\mathbf{q}\right).
	\end{split}
\end{equation}
Here, $s^{\eta}=+1 (-1)$ for $\eta =y$ (otherwise), $\tilde{\sigma}^\eta=\sigma^\eta$ for $\eta \in \{0,x,z\}$ and $\tilde{\sigma}^y=i\sigma^y$. The zeroth component is the unpolarized impurity DOS i.e. $\delta  S_{0}\rightarrow \delta\rho$. 
At $\mathbf{q}=0$, the quantities $\delta \rho(\mathbf{q}=0,E) \rightarrow \delta \rho(E)$ and $\delta  S_{\eta }(\mathbf{q}=0,E)\rightarrow \delta  S_{\eta }(E)$ correspond to the spatially averaged disorder-induced DOS and SPDOS, respectively. We will focus on the dependence of  $\delta \rho(E)$ and $\delta  S_{\eta }(E)$ as a function of energy and impurity strength to establish the formation of subgap states.
%


%
\section{Chern number\label{sec:Chern}}
\subsection{Results}
\begin{figure}[h]
	\begin{center}
		\begin{tikzpicture}
			\node at (0,0) {\includegraphics[width=9cm]{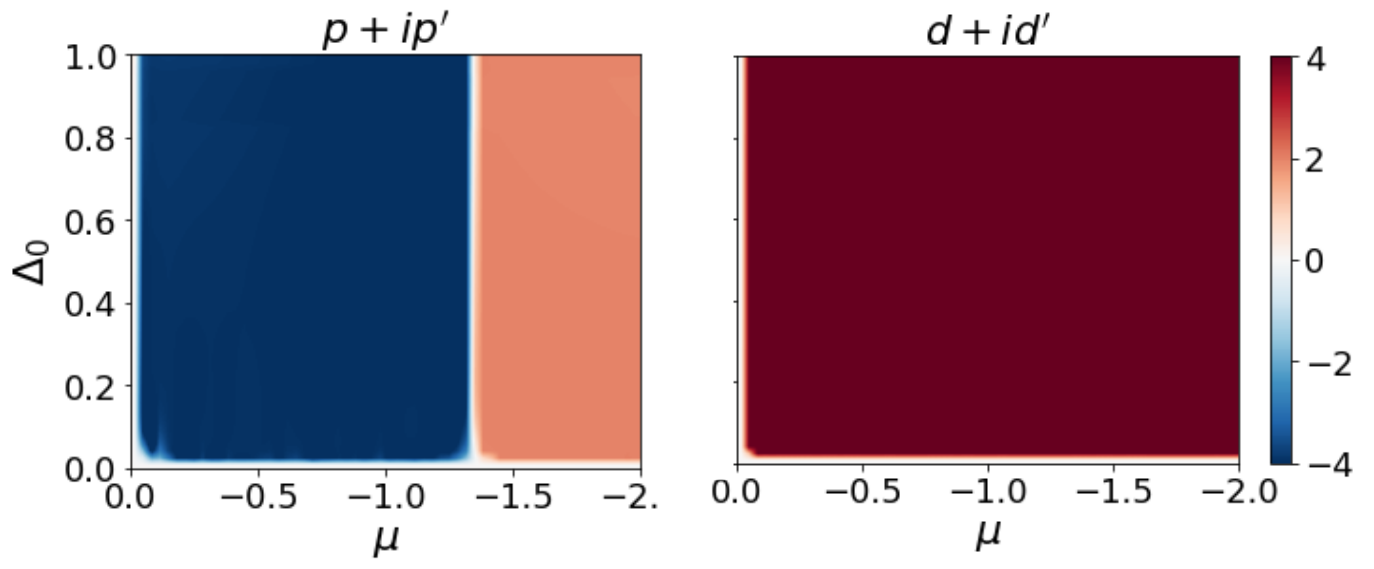}};
			\node at (-3.4,1.75) {(a)};
			\node at (0.8,1.75) {(b)};
		\end{tikzpicture}
	\end{center}
	\caption{Chern number as a function of the chemical potential $\mu$ and the superconducting
		coupling $\Delta_0$ for the 6-Band model with for $p+ip$ (a)- and
		$d+id'$ (b)- pairing symmetries.}
	\label{fig:Chern_6B1V} 
\end{figure}

In what follows we focus on calculating the  Chern number as a function of the chemical potential ($\mu$) and the SC pairing strength ($\Delta_0$). As only the chiral pairing states have a non-trivial Chern number, we will compute the Chern number only for the $p+ip'$ and $d+id'$ pairing symmetries in the absence of impurity. 
For a $d+id'$ pairing symmetry, see Fig.~\ref{fig:Chern_6B1V}(b), the Chern number takes a constant number of $4$ independently of the value of $\mu$ and $\Delta_0$. This is consistent with similar observations made for a SC monolayer graphene \cite{Adeline2022} up to a minus sign because $\mu$ is negative in the present study. This is also in agreement with Ref.~\onlinecite{wu2019topological}, and suggests that this is quite a generic feature of the $d+id'$ pairing.

For a $p+ip'$ pairing the Chern number is independent of $\Delta_0$ but shows a non-trivial $\mu$ dependence, see Fig.~\ref{fig:Chern_6B1V}(a). With our set of parameters defined in Table~\ref{tab:parameter}, Appendix A, the critical chemical potential is $\mu_{\text{critical}}\approx -1.3$ coincides with the point at which the FS becomes tangent to the boundary of the BZ, $\mu_T$, see Fig.~\ref{fig:FS_6B1V} (e). For $|\mu| \le |\mu_{\text{critical}}|$, the Chern number is equal to $-4$ and for $|\mu| \ge |\mu_{\text{critical}}|$ it is equal to $2$. This is consistent with the observation made for a superconducting moir\'eless graphene systems\cite{Adeline2022}, up to a minus sign. Note however that in regular graphene the topological transition occurs at the unique Lifshitz transition corresponding to a VHS in the DOS, while here it occurs at a point that has no special signature in the DOS. In the next subsection we will provide an intuitive explanation for this observation. This result shows however that for TBG, same as for moir\'eless superconducting graphene systems, the Fermi surface structure in the normal state dictates to a large extent its superconducting topological properties.

It should be noted that in the above analysis $\mathcal{C}$ is valley resolved since the 6B1V model contains only one of the degenerate valleys, as discussed in Section~\ref{sec:NormalModel}. Considering both valleys will double the $\mathcal{C}$ in Fig.~\ref{fig:Chern_6B1V}, so that one recovers as expected a total $\mathcal{C}$ similar to that of bilayer graphene, as calculated in Ref.~\onlinecite{Adeline2022}. Focusing on a single valley and a single isolated band, as prescribed by the 6B1V model, reduces the TBG physics to monolayer graphene-like.

For completeness and comparison, in Appendix \ref{app:2bandmodel} we also computed the Chern number for both $d+id'$ and $p+ip'$ order parameters for a 2-band model, which describes only the flat bands. As shown in Fig.~\ref{fig:Normal_state_band2}, the broad features of the two lowest energy bands of the 6B1V and the 2B1V model are very similar. This explains why we obtain similar phase diagrams for both models.

\subsection{Chern winding number formula}
To understand these results we study the winding formula\cite{kallin2016chiral},
equivalent to Eq.~(\ref{eq:TKNN}), which estimates the Chern number based on the winding of the order parameter around the Fermi surface. For one band the Chern number is given by:
\begin{flalign}
\mathcal{C}=\dfrac{1}{4\pi}\int_{BZ}\text{d}\mathbf{k}\hat{h}\cdot(\partial_{k_x}\hat{h}\times\partial_{k_y}\hat{h})
\label{eq:winding}
\end{flalign}
with $\mathbf{h}=(\text{Im}[\Delta(\mathbf{k})], \text{Re}[\Delta(\mathbf{k})],\epsilon(\mathbf{k})-\mu)$
where $\Delta(\mathbf{k})$ is the superconducting order parameter and $\hat{h}=\mathbf{h}/|\mathbf{h}|$. As this expression shows, we need $\mathbf{h}\ne \mathbf{0}$ or else this expression is undefined. It has been demonstrated that Eq.~(\ref{eq:winding}) is mathematically equivalent to integrating over an infinitesimal two-dimensional surface that encloses the special point where $\hat{h}$ is undefined \cite{yakovenko1990chern}, in the original literature such points were also denoted ``diabolic'' points. The value of $\mathcal{C}$ can only be altered by modifying the winding around these specific points. When the Fermi surface is far from these ``diabolic" points, any gradual deformation will be incapable of altering the value of $\mathcal{C}$, as one would expect for a topological invariant. Thus we expect that the only possible modifications of the Chern number may happen for TBG at $\mu_L=-0.63$ and $\mu_T=-1.3$, when the topology of the Fermi surface is changing and/or the FS is touching the diabolic points. In the following we note $D$ the set of diabolic points.



We can now explain the Chern results for both $p+ip'$ and $d+id'$ pairing symmetries. Using the form of the pairing symmetries, we have $D_{p+ip'}=\{\Gamma,K,M\}$ and $D_{d+id'}=\{\Gamma,K\}$, see Fig.~\ref{fig:FS_6B1V} where the $D_{p+ip'}$ points are denoted by red crosses and the $D_{d+id'}$ points are denoted by purple X's. Note that these points correspond to the nodes of the order parameters \cite{pangburn2022superconductivity}.


For the $d+id'$ state, since the $M$ point is not a special point, the only value of $\mu$ susceptible to generate a change in the Chern number is that corresponding to the VHS, when the FS changes from two contours winding around the K points to a single contour winding around the Gamma point. For $|\mu| \ge |\mu_L|$ the superconducting order parameter winds twice around the $\Gamma$-centered Fermi surface because the angular momentum is $l_{d+id'}=2$, leading to a Chern number of $4$ when adding the two spin contributions. For $|\mu| \le |\mu_L|$, same as for regular graphene\cite{Adeline2022}, each Fermi surface centered around the $K$ and $K^\prime$ contributes a (positive) unit to the Chern number per spin species, resulting in a $\mathcal{C} =4$ for the chiral $d+id'$-wave state also at $|\mu| \le |\mu_L|$. Thus the Chern number for the $d+id'$ is unaffected by this Lifshitz transition.

For the $p+ip'$ state, since $M$ is also a special point, there are two possible points at which the Chern number may change, at the  Lifshitz transition at $\mu_L=-0.63$ and when the FS is touching the BZ boundary at $\mu_T=-1.3$. The winding number of the $p+ip'$ pairing remains unchanged at the Lifshitz transition, and the underlying reason is quite subtle. Thus, below the Lifshitz transition, for $|\mu| \le |\mu_L|$, the $p+ip'$-wave order parameter winds once around both $K$ and $K'$ \cite{pangburn2022superconductivity}, thus giving a Chern number of $-2$ for each spin ($1$ per Fermi surface), hence a total Chern number of $-4$. On the other hand, for $|\mu| \ge |\mu_L|$, to calculate the winding number we need to consider the summation of windings around the $\Gamma$ point and around the three inequivalent $M$ points. Due to the fact that the winding around the $M$ point is the same but opposite in sign as that around the $\Gamma$ point, one of the $M$ points compensates for the winding around the $\Gamma$ point, and the Chern number $\mathcal{C}=-2\times(3-1)=-4$, same as in the phase where the Fermi surface encloses the two inequivalent $K$ points. On the other hand, at $\mu_T=-1.3$, the FS touches the $M$ points, and thus changes from winding around both the $M$ points and the $\Gamma$ point to winding solely around the $\Gamma$ point. Thus the total winding is reduced from two to one, and its sign also changes, yielding a total Chern of $2$ (one per each spin species).




%
\section{Impurity subgap states}\label{sec:Imp}
In what follows we investigate the subgap states induced by magnetic and non-magnetic (scalar) impurities. A recent study showed that for non-superconducting TBG, the flat bands are similarly affected irrespective of the region of the moir\'e pattern where an impurity is placed~\cite{baldo2023defectinduced}. In general, an impurity may induce scattering in all orbitals. However, the most relevant scattering are those that influence the low-energy levels around the Fermi energy.  As discussed in Sec.~\ref{sec:NormalFermiology}, these two bands originate mostly from the $p_\pm$-orbitals (except close to the $\Gamma$-point where the $p_z$-orbital contributes significantly). Motivated by these observations we consider an impurity affecting either the $p_z$- or the $p_\pm$-orbital, and three possible scenarios. The first one corresponds to the impurity acting solely on the $p_z$-orbital, the second involves an impurity acting on the $p_+$-orbital or $p_-$-orbital (same results is obtained for a $p_-$ one), and in the third one we consider impurities acting on the $(p_+,p_-)$-orbitals simultaneously. 

%
%
%
\subsection{Scalar impurity}\label{sec:LDOS}
\subsubsection{Impurity on a $p_z$ orbital}
\begin{figure*}[!htb]
	\includegraphics[width=4cm]{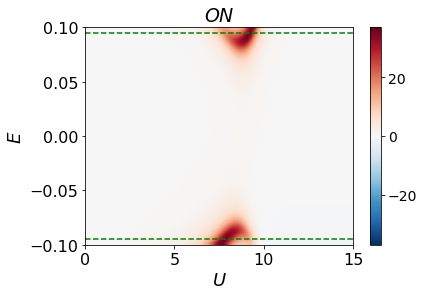} \includegraphics[width=4cm]{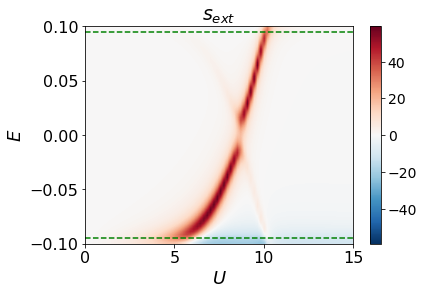}
	\includegraphics[width=4cm]{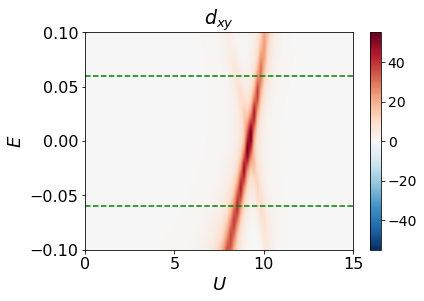} \includegraphics[width=4cm]{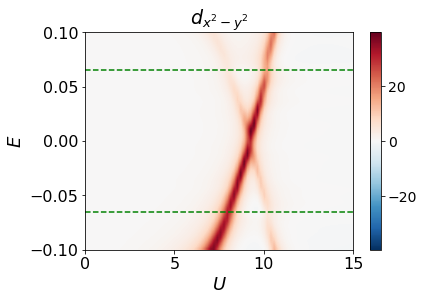}
	\vspace{-0cm}
	\includegraphics[width=4cm]{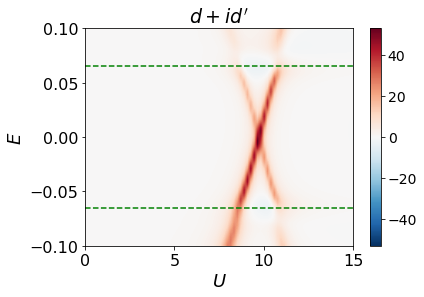} \includegraphics[width=4cm]{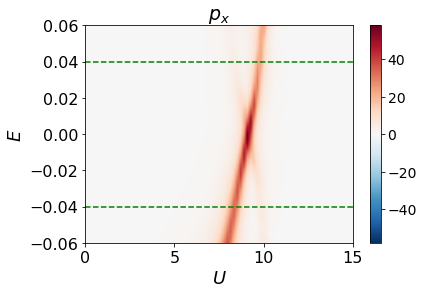}
	\includegraphics[width=4cm]{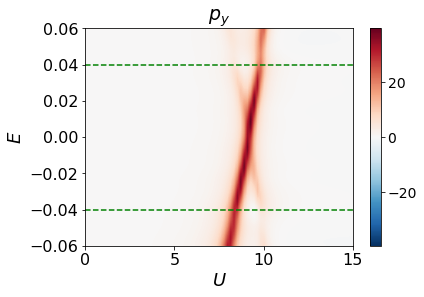} \includegraphics[width=4cm]{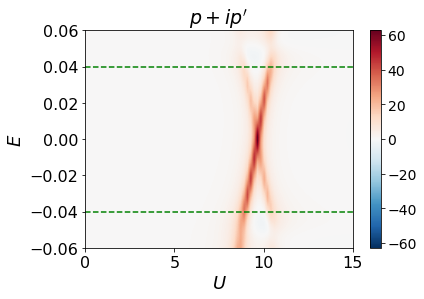}
	\caption{Impurity-induced DOS as a function of energy and impurity strength for scalar impurities acting on the $p_z$-orbital. The SC states considered are those with ON $s$-wave, NN $s_{\text{ext}}$-, $d_{xy}$-, $d_{x^{2}-y^{2}}$-,
		$p_{x}$-, $p_{y}$-, $p+ip\,'$-, $d+id\,'$-wave with $\Delta_0=0.1$ and $\mu=-0.4$. The gap is denoted
		by the dashed lines.}
	\label{fig:LDOS_scalar1} 
\end{figure*}
\begin{figure}[!htb]
\begin{tikzpicture}

	\node at (0,0) {\includegraphics[width=4.3cm]{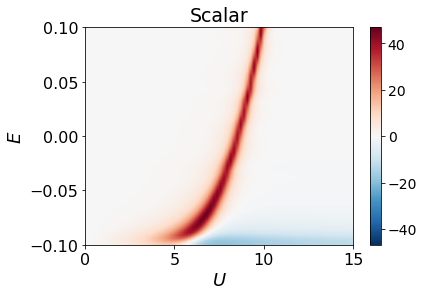}};
 \node at (4.5,0) 
	{\includegraphics[width=4.3cm]{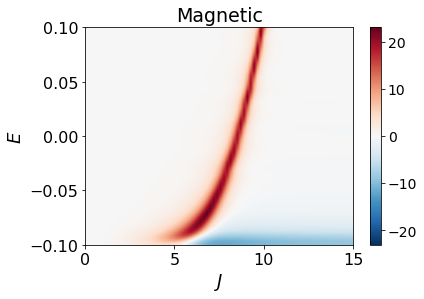}};
 \node at (-1.,1.4) {(a)};
			\node at (3.5,1.4) {(b)};
 \end{tikzpicture}
	\caption{Impurity-induced DOS as a function of energy and impurity strength for scalar (a) and magnetic (b) impurity acting on the $p_z$-orbital for nonsuperconducting TBG. The results are similar to the results of Fig.~\ref{fig:LDOS_scalar1}. As shown in Appendix.\ref{sec:Normal_state_pz}, the impurity-spin of the induced bound state is forced to be in the same direction as the spin of the impurity, which explains the reduction of the subgap intensity by a factor of 2 between the scalar and magnetic impurity.}
	\label{fig:LDOS_Normal} 
\end{figure}
%
%
%
When the impurity only acts on the $p_z$ orbitals, the subgap states are very similar between different pairing symmetries, except for the onsite $s$-wave, see Fig.~\ref{fig:LDOS_scalar1}.
Also, the impurity states behave almost in the same manner as in the normal state, showing an approximate inverse relationship with the impurity strength i.e $E \propto 1/ U$, known for the impurity state behavior in the normal state of Dirac systems~\cite{Wehling2014Dirac,Awoga2018}. This is because the $p_z$ orbital is at the origin not superconducting. In Fig.~\ref{fig:LDOS_Normal}, we show the LDOS resulting from normal-state impurities, confirming that the impurity states seen in Fig.~\ref{fig:LDOS_scalar1} are indeed mostly the result of normal state physics. The only exception is the onsite $s$-wave where the induced gap is large and completely wipes out the normal-state impurity physics.  

Nevertheless, due to the hybridization between the orbitals, some superconductivity is also induced indirectly in the $p_z$ orbital. Explicit derivations of this phenomenon are relegated to Appendices~\ref{sec:pz_orbital} and~\ref{sec:Symmetry_diagnosis}. The induced superconductivity influences the impurity states in two ways. 
First, the dispersion of the impurity states may be affected, note for example the slightly different slopes of the impurity states in Fig.~\ref{fig:LDOS_scalar1} compared to the normal state Fig.~\ref{fig:LDOS_Normal}. Such an effect has been reported in other graphene systems~\cite{Awoga2018,Haurie2022shiba}. Secondly, at the gap energy and below, the impurity states are significantly modified. 

\subsubsection{Impurity on a $p_+/p_-$ orbital}
\begin{figure*}[!htb]
	\includegraphics[width=4cm]{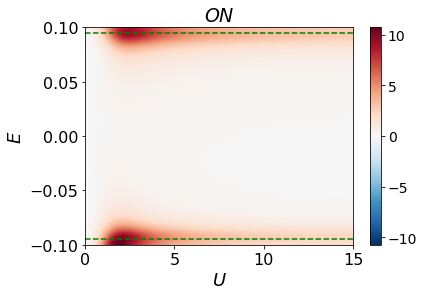} \includegraphics[width=4cm]{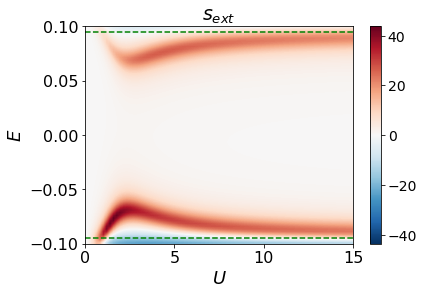}
	\includegraphics[width=4cm]{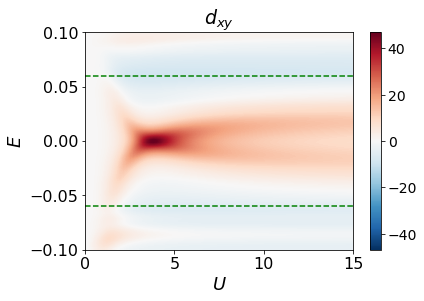} \includegraphics[width=4cm]{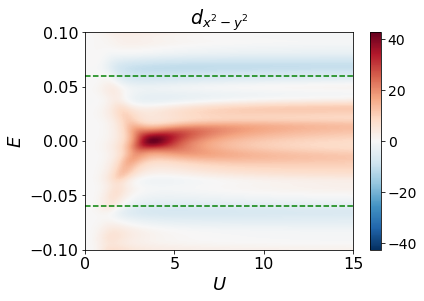}
	\vspace{-0cm}
	\includegraphics[width=4cm]{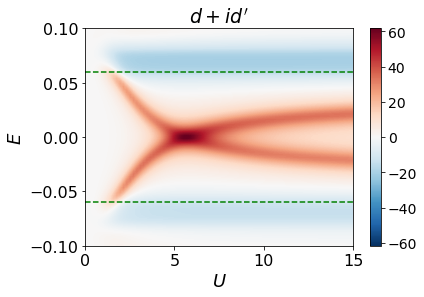} \includegraphics[width=4cm]{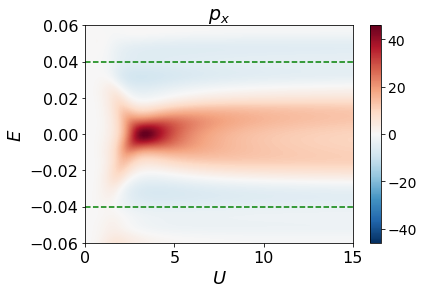}
	\includegraphics[width=4cm]{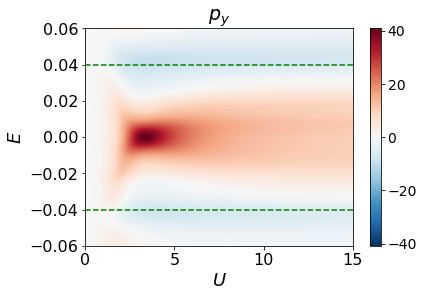} \includegraphics[width=4cm]{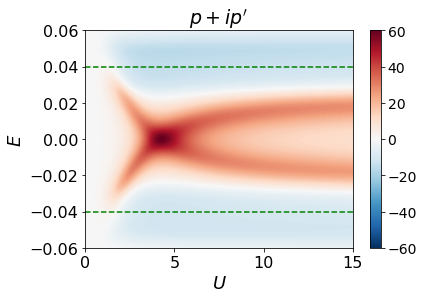}
	\caption{The figure shows the impurity-induced DOS as a function of energy and impurity strength for scalar impurities acting on the $(p_+)$ orbital. The SC states considered are those with ON $s$-wave, NN $s_{\text{ext}}$-, $d_{xy}$-, $d_{x^{2}-y^{2}}$-,
		$p_{x}$-, $p_{y}$-, $p+ip\,'$-, $d+id\,'$-wave with $\Delta_0=0.1$ and $\mu=-0.4$.  The gap is denoted
		by the dashed lines.}
	\label{fig:LDOS_scalar3} 
\end{figure*}

The effect of the scalar impurity on one of the superconducting orbitals $p_+$ or $p_-$ is similar to the case of superconducting graphene~\cite{Awoga2018,Haurie2022shiba}, see Fig.~\ref{fig:LDOS_scalar3}.
An important distinction that we note is between $s$-wave and all other pairing symmetries, such as $p$-wave and $d$-wave. The absence of subgap bound states for a scalar impurity in $s$-wave pairing is due to the fact that this pairing symmetry transforms trivially under point group transformations. As a consequence, the $T$-matrix of the $p_z$-orbital contains a constant term which prevents subgap states, see Appendix~\ref{sec:pz_orbital}. This result is consistent with the Anderson theorem which states that $s$-wave SCs are robust against scalar impurities \cite{Anderson1959TheoryOD}. In contrast, the $p$-wave and $d$-wave superconducting states exhibit two subgap resonance states that cross zero energy as expected~\cite{pan2000imaging,Awoga2018}.

\begin{figure}
    \centering
    \begin{tikzpicture}
        
   \node at (0,0){
\includegraphics[width=7cm]{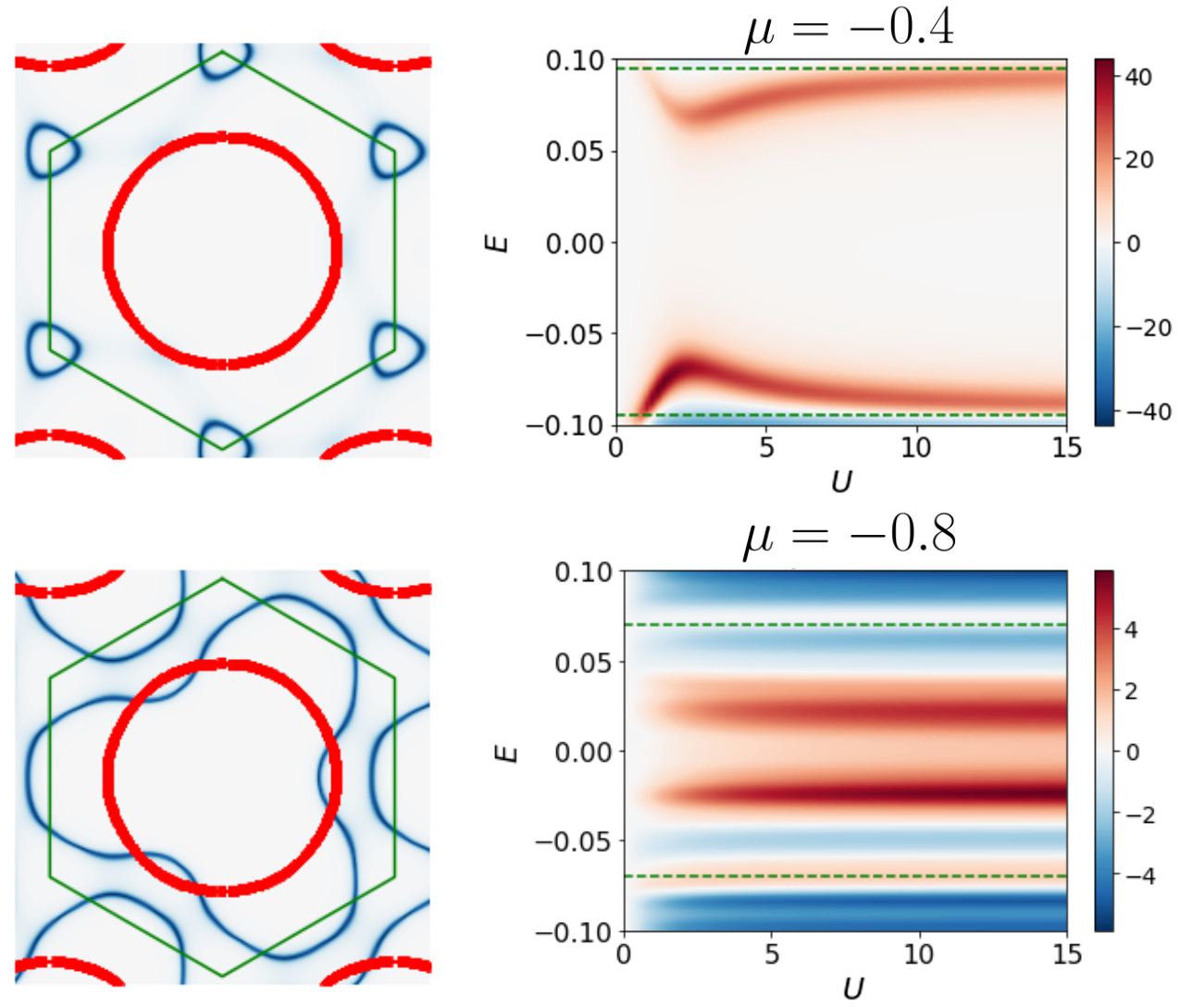}};
\node at (-3,2.8) {(a)};
\node at (0.5,2.8) {(b)};
\node at (-3,-0.2) {(c)};
\node at (0.5,-0.2) {(d)};
     \end{tikzpicture}
     \caption{Figures (a) and (c) show the Fermi surface (in blue) for $\mu=-0.4$ (a)  and  $\mu=-0.8$ (c), and the zeroes of $\Delta^{s_{ext}}(\mathbf{k})$ (in red). Figures (b) and (d) show the corresponding impurity-induced DOS as a function of energy and impurity strength for a scalar impurity acting on the $p_+$ orbital for an $s_{ext}$ pairing symmetry. We take $\Delta_0=1$. The gap is denoted by the green dashed lines.}
    \label{fig:sWaveAnderson}
\end{figure}
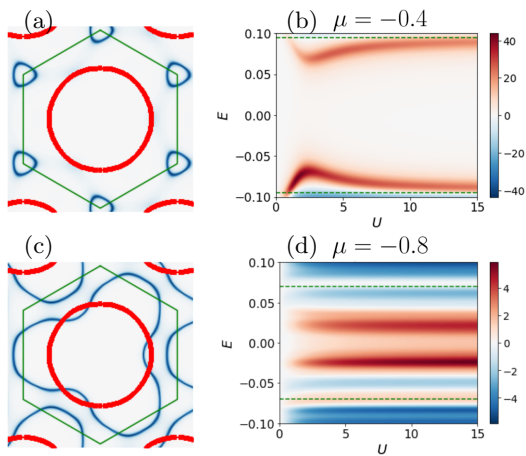

Note that the Anderson theorem provides understanding about global quantities, and thus only applies approximately to local ones such as the LDOS in the vicinity of the impurity. Thus it does not completely forbids the formation of subgap states, and here we do not see indeed a full suppression of subgap states in the $s$-wave superconducting state, but rather the formation of some states that are pushed towards the gap edges, that do not cross zero-energy and cannot be fully counted as subgap states. Note also that the Anderson theorem does not always necessarily apply to the $s_{ext}$-wave state, since the phase of the order parameter may vary across the Fermi surface. In Fig.~\ref{fig:sWaveAnderson}, we plot the FS and the subgap states for two values of the chemical potential. For the first one ($\mu=-0.4$), there is no intersection between the Fermi surface and the zeroes of the $s_{ext}$-wave order parameter. This means the phase of $\Delta(\mathbf{k})$ is approximately constant on the Fermi surface (see Fig.~\ref{fig:sWaveAnderson}(a)), and indeed, we only see two gap-edge approximate subgap states (Fig.~\ref{fig:sWaveAnderson}.(b)). For the second chemical potential ($\mu=-0.8$), there is is an intersection between the zeroes of $\Delta(\mathbf{k})$ and the Fermi surface (Fig.~\ref{fig:sWaveAnderson}(c)). In this case the Anderson theorem does not apply and we see the formation of two full subgap states (Fig.~\ref{fig:sWaveAnderson}(d)).

%
%
\subsubsection{Impurity on both $(p_+,p_-)$ orbitals}

\begin{figure*}[!htb]
\includegraphics[width=4cm]{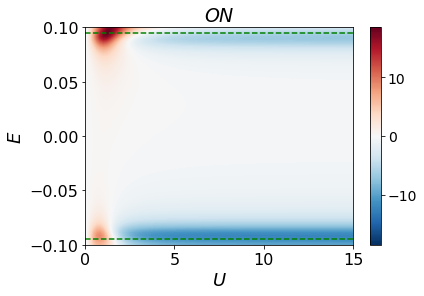} \includegraphics[width=4cm]{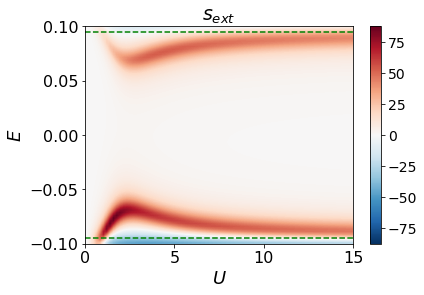}
\includegraphics[width=4cm]{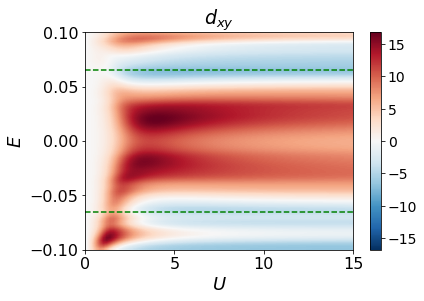} \includegraphics[width=4cm]{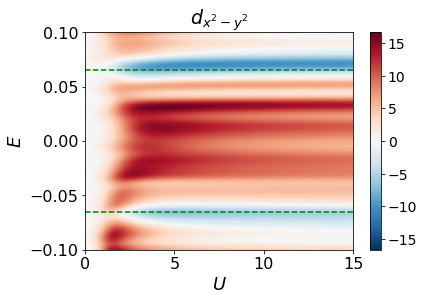}
\vspace{-0cm}
\includegraphics[width=4cm]{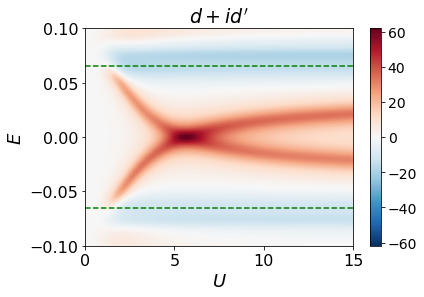} \includegraphics[width=4cm]{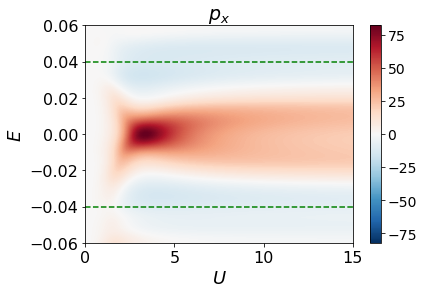}
\includegraphics[width=4cm]{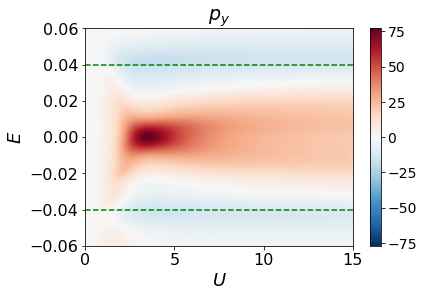} \includegraphics[width=4cm]{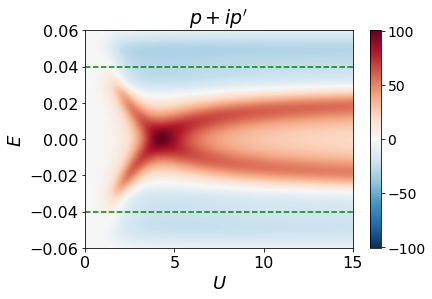}
\caption{Impurity-induced DOS as a function of energy and impurity strength for scalar impurities acting on the $(p_+,p_-)$ orbitals. The SC states considered are those with ON $s$-wave, NN $s_{\text{ext}}$-, $d_{xy}$-, $d_{x^{2}-y^{2}}$-,
$p_{x}$-, $p_{y}$-, $p+ip\,'$-, $d+id\,'$-wave with $\Delta_0=0.1$ and $\mu=-0.4$.  The gap is denoted
by the dashed lines.}
\label{fig:LDOS_scalar2} 
\end{figure*}

When impurities affect both superconducting orbitals $p_+$ and $p_-$ simultaneously, a more precise understanding of the pairing can be obtained by distinguishing between intra-orbital and inter-orbital coupling components for a given pairing, as explained in Appendix~\ref{sec:Number_states}. The subgap states for $s$-wave pairings remain unchanged between a single orbital impurity $p_+$(or $p_-$) and a combined $(p_+,p_-)$ impurity, as shown in Fig.~\ref{fig:LDOS_scalar2}, except for a doubling of the intensity due to the doubling of the number of states coming from the two distinct orbital impurities. This is consistent with the notion that the physics of these pairings is dominated by the intra-orbital component (see Appendix~\ref{sec:Number_states}).

In what concerns the pairings dominated by inter-orbital components (such as the the $d$-wave and $p$-wave), they can be separated between two classes : type 1 (which includes $d_{xy}$-, $d_{x^2-y^2}$-, $p_x$-, and $p_y$-wave pairing) and type 2 (which includes $d+id'$- and $p+ip'$-wave pairing). The origin of this distinction can be understood through a symmetry analysis of the $p_+$ and $p_-$ orbitals of the tight-binding model, as explained in Appendix~\ref{sec:symmetry_pwave}. 
For the type 1 pairings, having an impurity on both the $(p_+,p_-)$ orbitals strongly affect the  position of the poles of the $T$-matrix, and thus can reduce the number of subgap bound states obtained for a single $p_\pm$ impurity, depending on the specific details of the model and the pairing. Thus we see here that the subgap states for $d_{xy}$- and $d_{x^2-y^2}$-wave pairings disappear, leaving only heavily damped and ill-defined states within the gap. In contrast, the $p_x$- and $p_y$-wave pairings, while also dominated by the same type of inter-orbital effects, exhibit normal components of the Green's function that dominate over the anomalous ones. This explains why for these pairings the subgap bound states remain unchanged when the impurity affects both $(p_+,p_-)$ orbitals, except for a doubling of the intensity which is expected in this case (see Appendix~\ref{sec:Number_states}).

The type 2 pairings exhibit a distinct behavior compared to type 1 when subjected to the second impurity potential. For the $d+id'$ states, the subgap bound states that were present when the impurity potential acted on a single orbital are not affected by the second impurity potential, thus both the number and the intensity of the observed states is the same for a single and double orbital impurity. Note that this is valid only for scalar impurities and not too large inter-orbital couplings, and extra states will be indeed visible for magnetic impurities in the next section. However, the $p+ip'$ state, like the other triplet pairings, is also dominated by the normal components of the local Green function, which is why we observe degenerate impurity states when the impurity is located on both $(p_+,p_-)$ orbitals, and a doubling of the DOS intensity. 

Moreover, it is worth noting that for type 2 inter-orbital pairings, subgap bound states are guaranteed to exist even for arbitrarily high values of the impurity potential. This is in contrast to the case where the impurity acted only on the $p_z$ orbital, where the strength of the impurity potential had to be carefully tuned to observe subgap bound states. 

When impurities act on all three orbitals localized on a triangular lattice, the resulting effects can be approximated as the sum of the effects when only the $p_z$ and $(p_+,p_-)$ orbitals are introduced. However, in the case of non-isotropic pairing symmetry, such as $d$-wave, there is an additional coupling between these two impurity states. This coupling allows for hybridization between the modes associated with the $p_z$ orbitals and the $(p_+,p_-)$ orbitals, which can vary in strength depending on the microscopic details of the inter-orbital effects.

Table.~\ref{tab:summaryscalar} provides a summary of the number of states corresponding to each pairing symmetry for each scalar impurity acting on different sets of orbitals. For a scalar impurity, every bound states is spin degenerated. This degeneracy can be lifted by introducing a magnetic impurity. The numbers in Table.~\ref{tab:summaryscalar} correspond to a value for the chemical potential of $\mu=-0.4$, for which the FS consists of contours centered around the K points. 

%
%
\begin{table*}[!htb]
\begin{tabular}{|c|c|c|c|c|c|c|c|c|}
\hline 
Impurity  & ON & $s_{ext} $& $d_{xy} $& $d_{x^2-y^2} $& $d+id' $& $p_x $& $p_y $& $p+ip' $\\
\hline 
$p_z$ & 0 & $2^\star$ & 2 & 2 & 2 & 2 & 2 & 2\\
$p_+/p_-$ & 0 & $2^\star$ & 2 & 2 & 2 & 2 & 2 & 2\\
$(p_+,p_-)$ & 0 & $2^\star$ (2) & 0 & 0 & 2 & 2 (2) & 2 (2) & 2 (2)\\
Coupling & intra & intra & inter-1 & inter-1 & inter-2 & inter-1 & inter-1 & inter-2 \\
$(p_+,p_-,p_z)$ & 4 & 4 & 4 & 4 & 8 & 4 & 4 & 2 (1) + 2(2) \\
\hline
\end{tabular}
\caption{Number of states per symmetry for a scalar impurity depending on which orbitals on the triangular lattice are affected by the impurity. We take $\mu=-0.4$. The predominant type of coupling for each pairing symmetry, which can be either intra-orbital (intra) or inter-orbital of type 1 (inter-1) or 2 (inter-2), is also listed. The symbol $\star$ is used to indicate instances where the subgap bound states are ``approximate'' gap-edge states and not ``full'' subgap states. In parenthesis we indicate the orbital degeneracy of the subgap states.}
\label{tab:summaryscalar}
\end{table*}
%

%
\subsection{Magnetic impurity\label{sec:SPLDOS}}
\begin{figure*}[!htb]
\includegraphics[width=4cm]{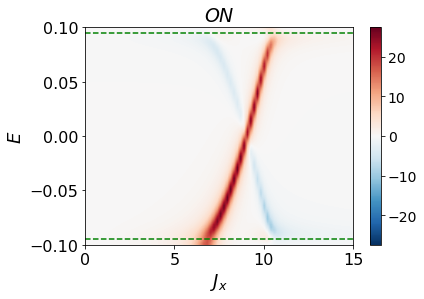} \includegraphics[width=4cm]{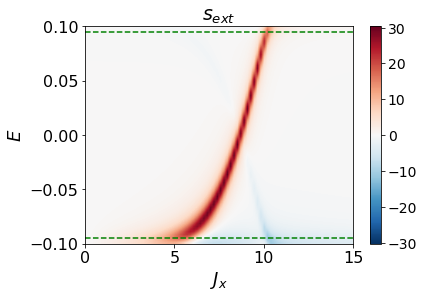}
\includegraphics[width=4cm]{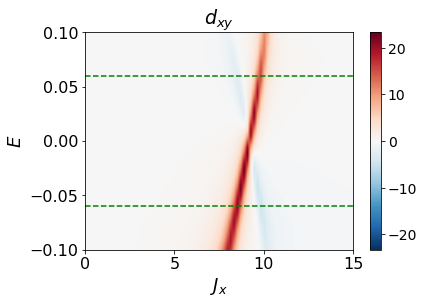} \includegraphics[width=4cm]{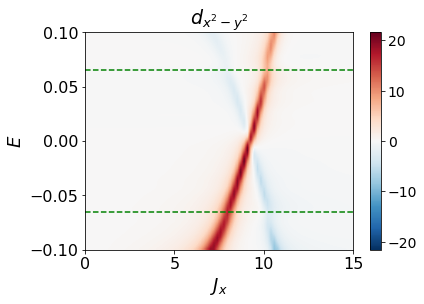}
\vspace{-0cm}
\includegraphics[width=4cm]{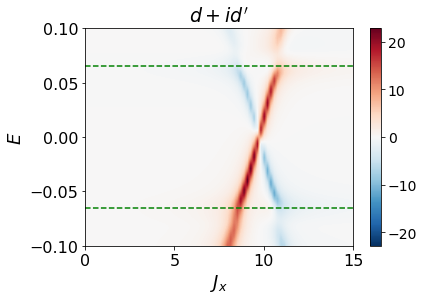} \includegraphics[width=4cm]{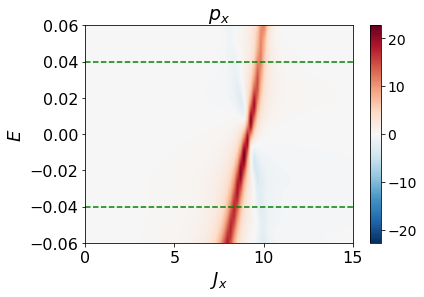}
\includegraphics[width=4cm]{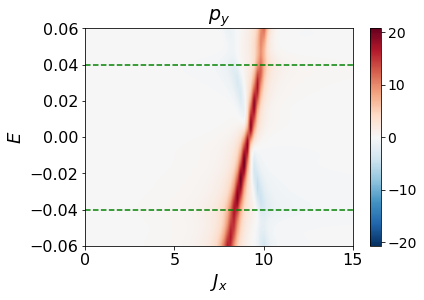} \includegraphics[width=4cm]{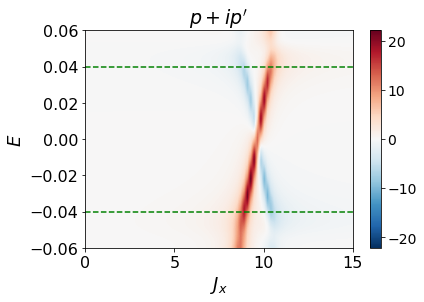}
\caption{Impurity-induced SPDOS as a function of energy and impurity strength for magnetic impurities with spin along x and acting on the $(p_z)$ orbitals. The superconducting states considered are those with ON $s$-wave, NN $s_{\text{ext}}$-, $d_{xy}$-, $d_{x^{2}-y^{2}}$-,
$p_{x}$-, $p_{y}$-, $p+ip\,'$-, $d+id\,'$-wave with $\Delta_0=0.1$ and $\mu=-0.4$.  The gap is denoted
by the dashed lines.}
\label{fig:SPDOS_x1} 
\end{figure*}
\begin{figure*}[!htb]
\includegraphics[width=4cm]{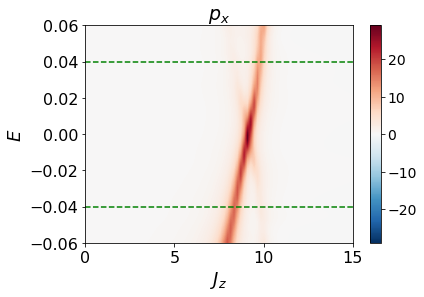}
\includegraphics[width=4cm]{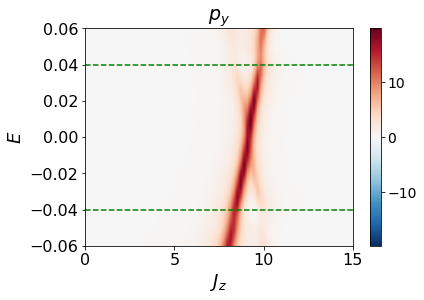} \includegraphics[width=4cm]{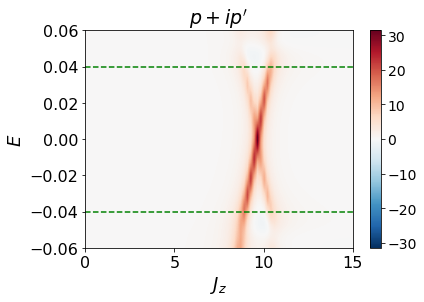}
\caption{Impurity-induced SPDOS as a function of energy and impurity strength for magnetic impurities with spin along z and acting on the $(p_z)$ orbitals. The superconducting states considered are those with nearest-neighbor $p_{x}$-, $p_{y}$-, $p+ip^\prime$-wave pairing, with $\mu=-0.4$ and $\Delta_{0}=0.1$. The gap is indicated by the dashed lines.}
\label{fig:SPDOS_z1} 
\end{figure*}

Time-reversal symmetry (TRS) is broken by magnetic impurities, which have a non-trivial spin dependence. As a result, magnetic impurities act as pair breakers even if $\mathbf{k}$ is constant on the Fermi surface. As shown in Figs.~\ref{fig:SPDOS_x1},~\ref{fig:SPDOS_x2},~\ref{fig:SPDOS_x3}, sub-gap states are obtained for all symmetries and all possible choices of orbitals acted upon by the impurity, including the on-site s-wave pairing which did not show subgap states in the presence of a scalar impurity.

\begin{figure*}[!htb]
\includegraphics[width=4cm]{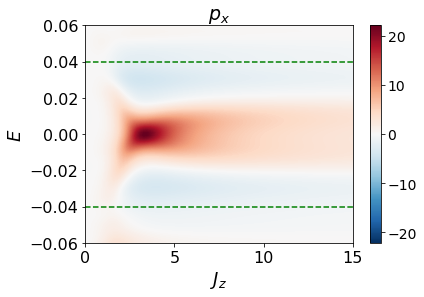}
\includegraphics[width=4cm]{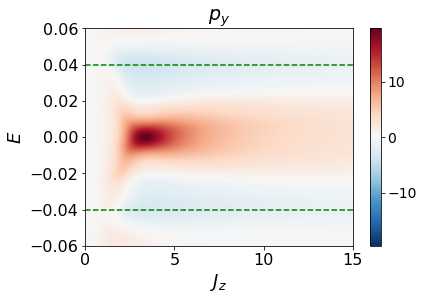} \includegraphics[width=4cm]{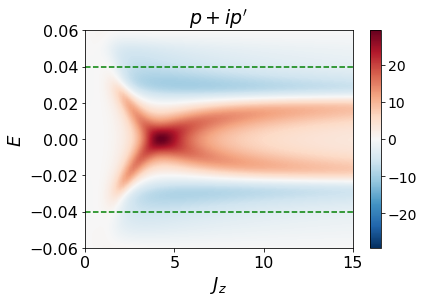}
\caption{Impurity-induced SPDOS as a function of energy and impurity strength for magnetic impurities with spin along z and acting on the $(p_+/p_-)$ orbital. The superconducting states considered are those with nearest-neighbor $p_{x}$-, $p_{y}$-, $p+ip^\prime$-wave pairing with $\mu=-0.4$ and $\Delta_{0}=0.1$. The gap is indicated by the dashed line.}
\label{fig:SPDOS_z3} 
\end{figure*}

\begin{figure*}[!htb]
\includegraphics[width=4cm]{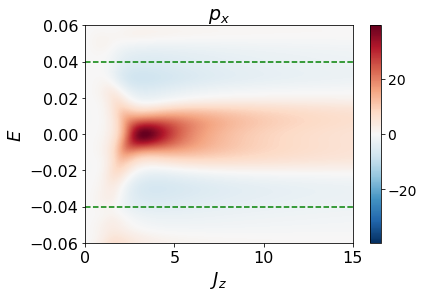}
\includegraphics[width=4cm]{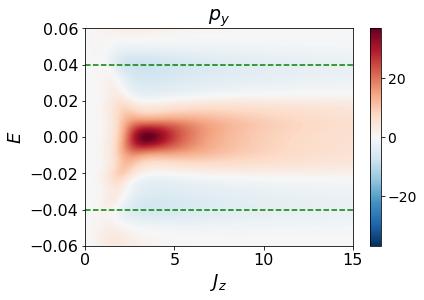} \includegraphics[width=4cm]{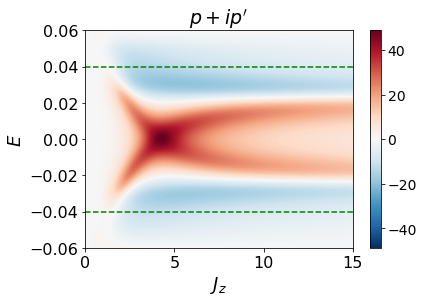}
\caption{Impurity-induced SPDOS as a function of energy and impurity strength for magnetic impurities with spin along z and acting on the $(p_+,p_-)$ orbitals. The superconducting states considered are those with nearest-neighbor $p_{x}$-, $p_{y}$-, $p+ip^\prime$-wave pairing, with $\mu=-0.4$ and $\Delta_{0}=0.1$. The gap is indicated by the dashed lines.}
\label{fig:SPDOS_z2} 
\end{figure*}

The SPDOS in the presence of a magnetic impurity can distinguish between spin singlet and spin triplet superconductors, as already demonstrated in Ref.~\onlinecite{Haurie2022shiba}. For singlet order parameters, the SPDOS shows the same qualitative behavior regardless of the direction of the impurity spin. However, for triplet order parameters, due to the non-trivial spin structure, the direction of the impurity spin matters. Thus, we compute the SPDOS for an impurity spin aligned with the Cooper-pair spin ($x$-direction), as well as for an orthogonal one ($y/z$-direction), as shown in Figs.~\ref{fig:SPDOS_z1},~\ref{fig:SPDOS_z3} and~\ref{fig:SPDOS_z2}. 

Our main observations are that for an impurity spin along the $x$-direction the opposite-energy states always have opposite spin polarization.
For the $y$- and $z$-directions, the two opposite-energy subgap states have the same spin, resulting in a non-zero average spin, as shown for example in Fig.~\ref{fig:SPDOS_z2}. An analytical derivation of the characteristics of the spin polarization of the impurity states for both spin-singlet and spin-triplet pairings is presented in detail in Appendix~\ref{sec:spin_polarization}. Furthermore, we find that the dependence of the spin polarization on the chemical potential $\mu$ is non-trivial and varies differently, depending on the type of orbital that the impurity is acting on, as shown in Fig.~\ref{fig:SPDOS_mu_pip} and discussed in Appendix~\ref{sec:FS_topology}. We will now present our observations in more detail.

%
\subsubsection{Impurity on a $p_z$ orbital}
Since the $p_z$-orbitals are not superconducting, this configuration does not differentiate between various gapped and gapless order parameter symmetries, and generally shows normal state behavior at energies above the gap, similar to Fig.~\ref{fig:LDOS_Normal}. In Appendix~\ref{sec:pz_orbital}, we demonstrate analytically that there are only two particle-hole symmetric subgap states that can exist in all cases. Fig.~\ref{fig:SPDOS_x1} shows these states for the various order parameter symmetries.

The SPDOS for spin-triplet order parameter and for an impurity spin in the y/z-directions exhibits a non-zero average spin polarization, as shown in Fig.~\ref{fig:SPDOS_z1}. 
The behavior of the impurity states with $\mu$ is discussed in more detail in Section ~\ref{sec:ImpChem}.


\subsubsection{Impurity on a $p_+/p_-$ orbital}
\begin{figure*}[!htb]
\includegraphics[width=4cm]{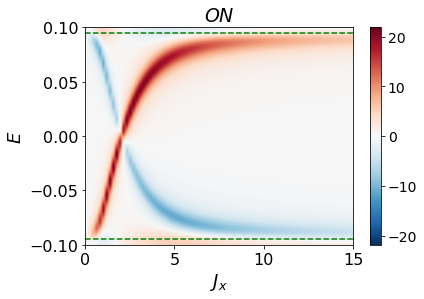} \includegraphics[width=4cm]{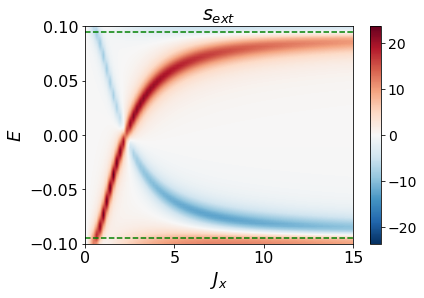}
\includegraphics[width=4cm]{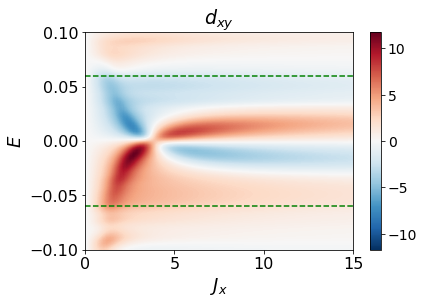} \includegraphics[width=4cm]{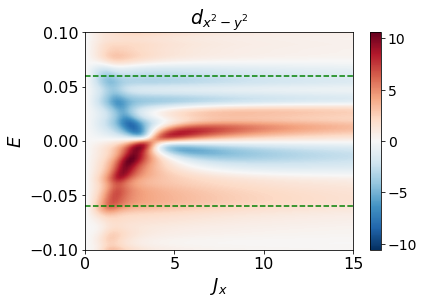}
\vspace{-0cm}
\includegraphics[width=4cm]{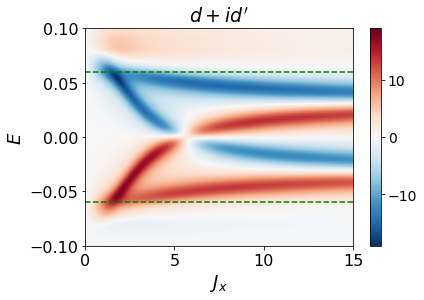} \includegraphics[width=4cm]{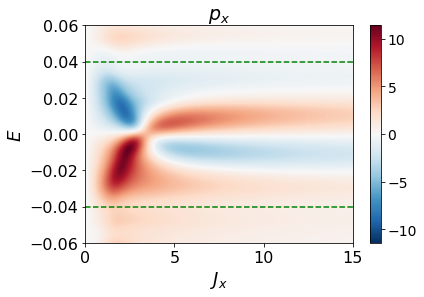}
\includegraphics[width=4cm]{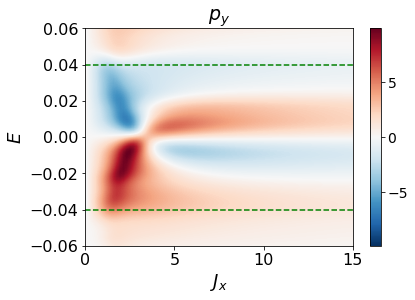} \includegraphics[width=4cm]{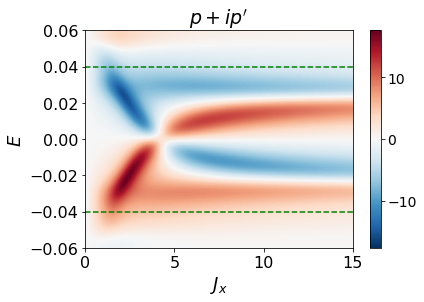}
\caption{Impurity-induced SPDOS as a function of energy and impurity strength for magnetic impurities with spin along x and acting on the $(p_+/p_-)$ orbital. The superconducting states considered are those with ON $s$-wave, NN $s_{\text{ext}}$-, $d_{xy}$-, $d_{x^{2}-y^{2}}$-,
$p_{x}$-, $p_{y}$-, $p+ip\,'$-, $d+id\,'$-wave with $\Delta_0=0.1$ and $\mu=-0.4$. The gap is denoted
by the dashed lines.}
\label{fig:SPDOS_x3} 
\end{figure*}

The action of a magnetic impurity on the superconducting orbitals can reveal more information about the pairing symmetries. Let's first consider the case where the impurity acts on only one of the superconducting orbitals. As $p_+$ and $p_-$ are degenerate, it does not matter which one the impurity acts on. Depending on the pairing symmetry, either the intra- or inter-orbital physics can dominate. Regardless of the specific microscopic details of the pairing mechanism, this leads to a prediction for the number of subgap states that we should expect to observe, as detailed in Appendix~\ref{sec:ppm_orbitals}. The general principle is that if intra-orbital physics dominates there should be two bound states, whereas if inter-orbital physics dominates there should be four bound states. The results shown in Fig.~\ref{fig:SPDOS_x3} allow us to conclude, as expected from the symmetry analysis, see Appendix~\ref{sec:Symmetry_diagnosis}, that ON and $s_{\text{ext}}$-wave pairing are dominated by intra-orbital physics, whereas $p$-wave and $d$-wave pairing are dominated by the inter-orbital components. The number of subgap bound states for the gapped pairing states, such as $p+ip'$ and $d+id'$, is well-defined and can be easily determined. However, for the nodal order parameters such as $d_{xy}$, $d_{x^2-y^2}$, $p_x$, and $p_y$, the situation is less clear. In these cases, the quasiparticles inside the gap are damped, resulting in virtual bound states that may not always be visible. Although we cannot always see these states with the set of parameters chosen in Fig.~\ref{fig:SPDOS_x3}, we have numerically confirmed that one can have a maximum of four states for the gapless $d$- and $p$-wave order parameters.
%
\subsubsection{Impurity on both $(p_+,p_-)$ orbitals\label{sec:Magnetic_ppm}}

\begin{figure*}[!htb]
\includegraphics[width=4cm]{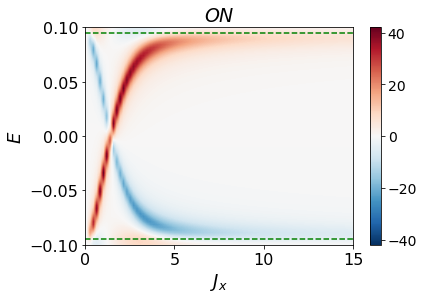} \includegraphics[width=4cm]{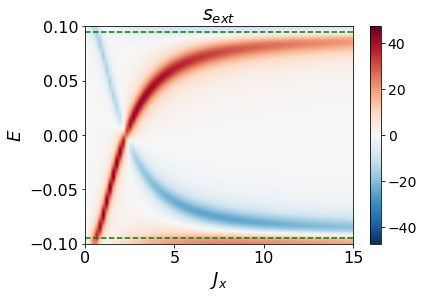}
\includegraphics[width=4cm]{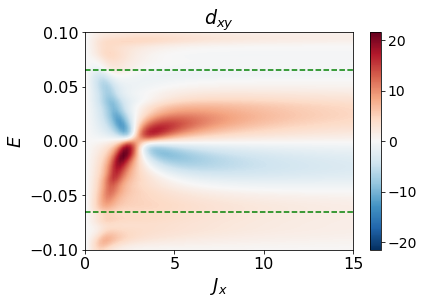} \includegraphics[width=4cm]{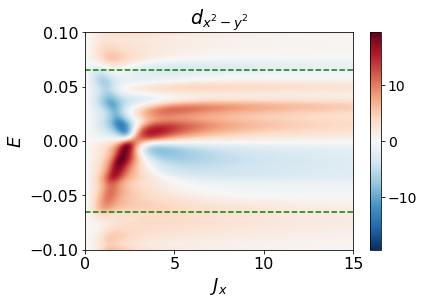}
\vspace{-0cm}
\includegraphics[width=4cm]{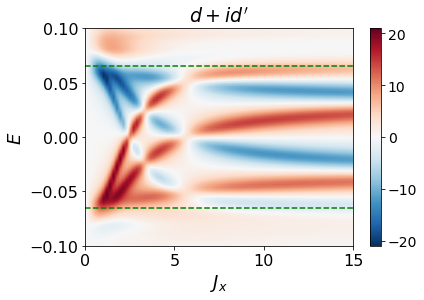} \includegraphics[width=4cm]{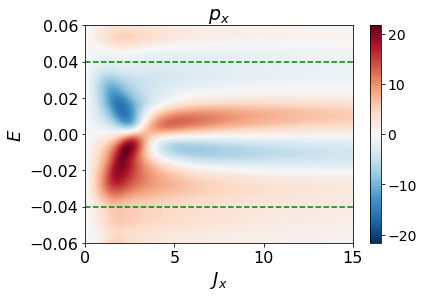}
\includegraphics[width=4cm]{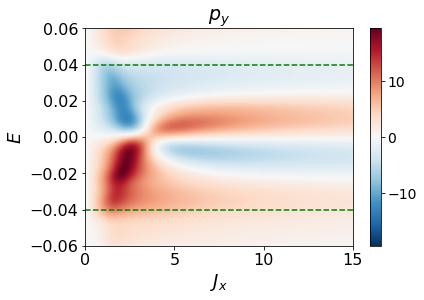} \includegraphics[width=4cm]{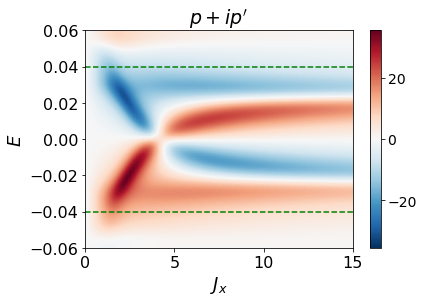}
\caption{Impurity-induced SPDOS as a function of energy and impurity strength for magnetic impurities with spin along x and acting on the $(p_+,p_-)$ orbitals. The superconducting states considered are those with ON $s$-wave, NN $s_{\text{ext}}$-, $d_{xy}$-, $d_{x^{2}-y^{2}}$-,
$p_{x}$-, $p_{y}$-, $p+ip\,'$-, $d+id\,'$-wave with $\Delta_0=0.1$ and $\mu=-0.4$.  The gap is denoted
by the dashed lines.}
\label{fig:SPDOS_x2} 
\end{figure*}


Even more information on the pairing symmetry can be obtained if the impurity acts on both superconducting orbitals $(p_+,p_-)$. A number of facts can be shown analytically in Appendix~\ref{sec:ppm_orbitals} and confirmed numerically in Fig.~\ref{fig:SPDOS_x2}. Thus, the subgap bound states for $s$-wave pairing, which is controlled by intra-orbital physics, are unchanged by the addition of a second impurity potential on the other superconducting orbital, except for a doubling of the intensity in the spectrum, same as for the scalar impurity case.

On the other hand, for $p$- and $d$-wave pairings, inter-orbital effects play a dominant role and we expect a different behavior for a single-orbital $p_\pm$ impurity and a  $(p_+,p_-)$ impurity. 
Thus, the second impurity potential has a significant impact on the type 1 pairings $d_{xy}$ and $d_{x^2-y^2}$, and can reduce the number of subgap bound states from 4 to 2 depending on the specific details of the model and the pairing. However, the analysis is complicated because all the pairings in this class are gapless, which requires considering a larger imaginary part that could dampen the in-gap bound states. We have examined different sets of parameters and observed generically that when we can observe four subgap bound states when the impurity potential acts only on one orbital, two of these four bound states vanish when the impurity acts on both orbitals.

On the other hand, for the $p$-wave type 1 pairing states $p_x$ and $p_y$, we find that adding a second orbital component to the impurity only doubles the intensity of the measured spectrum and does not significantly alter the number of the subgap bound states, same as in the case of a scalar impurity. As before this suggests that the normal components of the local Green function dominate over the anomalous ones. Note that this observation is not dictated by symmetries and may depend on the chosen set of parameter values.

The type 2 pairings exhibit a distinct behavior compared to type 1 when subjected to a combined $(p_+,p_-)$ impurity versus a single $p_\pm$ impurity. Thus, for the $p+ip'$ state, due to the small inter-orbital effects relative to the normal part of the Green function, one obtains two pairs of degenerate bound states, same as for a scalar impurity, and thus a doubling of the intensity for the $(p_+,p_-)$ impurity versus the $p_\pm$ impurity. On the other hand, for the $d+id'$ states, the subgap bound states that were present when the impurity potential acted on a single orbital are completely unaffected by the second impurity potential, nevertheless, quite spectacularly, an extra pair of states is formed, giving rise to six subgap bound states. 

Table~\ref{tab:summary} provides a summary of the number of states corresponding to each pairing symmetry for each magnetic impurity acting on different sets of orbitals. Once more we consider a value of the chemical potential, $\mu=-0.4$, corresponding to a FS which consists of contours centered around the K points. However, this number may change with increasing the chemical potential, since, as explained in Appendix \ref{sec:Number_states}, the count of subgap states for a specific pairing symmetry relies on the existence of a zero in the local Green function $G_e(E, \mathbf{R} = \mathbf{0})$ within the superconducting gap. The presence of such a zero can be altered at the topological phase transition occurring around $\mu \approx -1.3$, leading to a reduction in the number of subgap states, as depicted in Fig.~\ref{fig:SPDOS_mu_pip}. However this does not seem to be a generic feature, and we do not observe if for example in regular graphene.

%
%
\begin{table*}[!htb]
\begin{tabular}{|c|c|c|c|c|c|c|c|c|}
\hline 
Impurity  & ON & $s_{ext} $& $d_{xy} $& $d_{x^2-y^2} $& $d+id' $& $p_x $& $p_y $& $p+ip' $\\
\hline 
$p_z$ & 2 & 2 & 2 & 2 & 2 & 2 & 2 & 2\\
$p_+/p_-$ & 2 & 2 & 4/2 & 4/2 & 4 & 4/2 & 4/2 & 4\\
$(p_+,p_-)$ & 2 (2) & 2 (2) & 2 & 2 & 6 & 2 (2) & 2 (2) & 4 (2)\\
Coupling & intra & intra & inter-1 & inter-1 & inter-2 & inter-1 & inter-1 & inter-2 \\
$(p_+,p_-,p_z)$ & 4 & 4 & 4 & 4 & 8 & 4 & 4 & 2 (1)+2(2) \\
\hline
\end{tabular}
\caption{Number of states per symmetry for a magnetic impurity depending on which orbitals on the triangular lattice are affected by the impurity. We take $\mu=-0.4$. The predominant type of coupling for each pairing symmetry, which can be either intra-orbital (intra) or inter-orbital of type 1 (inter-1) or 2 (inter-2), is also listed. In parenthesis we indicate the orbital degeneracy of the subgap states when it is different from 1. The symbol $n/m$ indicates that the number of subgap states is generically $n$ but in some situations, due to the gapless nature of the order parameter, it may be reduced to $m$ visible states depending on the specific values of the parameters.}
\label{tab:summary}
\end{table*}

Before moving on we briefly discuss the scenario when the impurity is active in all $p$-orbitals. When impurities simultaneously act on all three orbitals localized on a triangular lattice, we observe results similar to those obtained in the case of scalar impurities. Specifically, for $s$-wave pairing, we find distinct sets of subgap bound states associated with each orbital, indicating minimal interaction between them. However, in the case of non-isotropic pairing, there can be a more pronounced hybridization between these two sets of subgap bound states, depending on the strength of inter-orbital effects.
\subsection{Chemical potential evolution of impurity states}\label{sec:ImpChem}
Hitherto, we have focused on impurity states at a fixed chemical potential, $\mu=-0.4$. In what follows we study the evolution of the impurity states for a chiral $p+ip^\prime$ SC TBG system as a function of $\mu$, for a $z$-direction impurity (which is the same as a $y$-direction impurity) and for different orbitals. We note that our results can be understood based on the the normal state topology, see Fig.~\ref{fig:FS_6B1V}.


%
\begin{figure*}[!htb]
	\includegraphics[width=4cm]{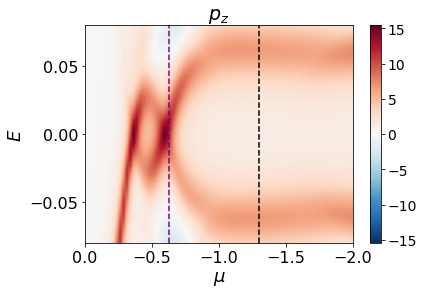}
	\includegraphics[width=4cm]{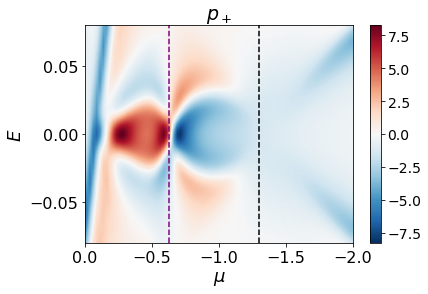} \includegraphics[width=4cm]{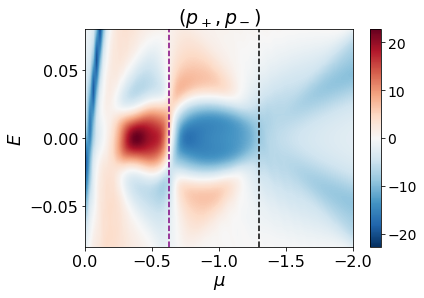}
	\caption{Impurity-induced $S_z(E)$ as a function of energy $E$ and the chemical potential $\mu$, for a $p+ip'$ state with impurities acting respectively on $(p_z)$ (a), $(p_+)$ (b), and $(p_+,p_-)$ (c) orbitals, with $J_z=10$ (a) and $J_z=4.5$ (b) and (c). We use $\Delta_0=0.1$. The chemical potential at which the Chern number is changing, $\mu\approx -1.3$, corresponds to the point where the normal state FS is touching the BZ boundary, and is denoted by the vertical black dashed line. The spin of the impurity states (for a $p_+$ and  $(p_+,p_-)$ impurity) flips at $\mu\approx -0.63$ (vertical purple dashed line), corresponding to the VHS and a change in FS topology from a contour centered around $\Gamma$ to contours centered around the K points.}
	\label{fig:SPDOS_mu_pip}
\end{figure*}
%

The most important observation is that for an impurity acting either on $p_+$ or on $(p_+,p_-)$, the triplet superconducting states show a non-trivial behavior, see Fig.~\ref{fig:SPDOS_mu_pip}(b,c).  For $0<|\mu| < 0.63$, there are 4 subgap states, and remarkably enough, for $\mu \approx -0.63$, close to the VHS and the Lifhitz transition, there is a flip of the spin for all the impurity states. This is a novel result which shows that, apart from the electron-hole conversion observed in the vicinity of the Lifshitz transition~\cite{Kim2016Charge,cao2016Superlattice}, there is a spin flip across this point as well. Furthermore, this provides a potential experimental evidence, accessible in spin-polarized STM experiment, for distinguishing between spin-triplet and spin-singlet superconducting states since the behavior is exclusive to spin-triplet superconductivity. 

Note also that at $\mu=-1.3$, marked by the dashed black line in Fig.~\ref{fig:SPDOS_mu_pip}, corresponding to the point at which one sees a topological phase transition and a change in the Chern number, for a $p_+$ or a $(p_+,p_-)$ impurity the number of states goes down from 4 to 2, and the energy of the Shiba states goes to zero. This is not the case for a $p_z$ impurity, see Fig.~\ref{fig:SPDOS_mu_pip}a). However, while this has also been seen in other superconduting systems such as in Ref.~\onlinecite{wang2012impurity,mashkoori2019impact} this is not a generic feature, as it has not been observed for example in regular graphene, and thus we cannot at this point claim that a change in the number of Shiba states may be generally used as an indication for a topological phase transition.

\section{Conclusion\label{sec:conclusion}}
We have computed the Chern number for TBG at the magic angle for all possible SC order parameters with low angular momentum. We have found a topological phase transition for the $p+ip'$ state at a chemical potential corresponding to the point where the Fermi surface is crossing the boundary of the first BZ. For TBG this point is different from the point where the FS changes topology from a contour centered around the Gamma points to contours centered around the K points, the latter being associated with a Van Hove singularity in the DOS. While for regular graphene systems the two points coincide and the topological phase transition occurs at the van Hove singularity, for TBG this occurs at a point that has no peculiar signature in the DOS. We have also found  a constant non-zero Chern number for the $d+id'$ state in TBG.

Our study has also focused on the effects of scalar and magnetic impurities on the DOS and the SPDOS. 
We propose that a spin-polarized scanning tunneling microscopy measurement can distinguish between spin-triplet and spin-singlet states in TBG, since spin-triplet states may exhibit opposite-energy impurity states with the same spin, while for the spin-singlet ones the opposite-energy Shiba states always have opposite spin. Moreover, for impurities interacting with the superconducting orbitals of the flat bands, the spin polarization of the spin-triplet states can flip at the twist-induced VHS.

Our results indicate that the subgap state formation in TBG is similar to what has been observed in monolayer and multilayer graphene \cite{Haurie2022shiba} for impurities localized in one of the $p_+$ or $p_-$ orbitals. However, because TBG is described by a multi-orbital model, the competition between intra- and inter-orbital effects leads to a much richer physics. Inter-orbital effects can be more or less important depending on the location of the impurity, as well as on the symmetries of the superconducting order parameter. 


\acknowledgments
OA acknowledges funding from NanoLund. MA acknowledges support from Spanish MICINN through the grant FIS2017-84860-R.

%

\appendix

\onecolumngrid
\section{6-Band model Hamiltonian\label{6B1Vmodel}}
In the basis defined in Eq.~\eqref{eq:Basis_6B1V} the Bloch Hamiltonian
is given by 
\begin{equation}
\hat{H}_{0}(\mathbf{k})=\begin{pmatrix}H_{p_{z}}+\mu_{p_{z}} & \hat{C}_{p_{\pm}p_{z}}^{\dagger} & \hat{0}\\
\hat{C}_{p_{\pm}p_{z}} & \hat{H}_{p_{\pm}}+\hat{\mu}_{p_{\pm}} & \hat{C}_{\kappa p_{\pm}}^{\dagger}\\
\hat{0} & \hat{C}_{\kappa p_{\pm}} & \hat{H}_{\kappa}+\hat{\mu}_{\kappa}
\end{pmatrix}\label{Eq:H0_6B1V}
\end{equation}
We denote all contributions to $\hat{H}_{0}(\mathbf{k})$ according
to their orbital character. We recall that the hat symbol indicates a quantity with a matrix structure in orbital space. We take $\phi_{lm}\equiv e^{-i\mathbf{k}.(l\mathbf{a}_{1}+m\mathbf{a}_{2})}$,
with $\mathbf{a}_{1}$ and $\mathbf{a}_{2}$ defined in Fig.\ref{fig:Lattice_6B1V}
and $\bar{l}\equiv-l$.  

First we have the nearest-neighbor hopping between the $(\tau,p_{z})$
orbitals 
\begin{equation}
H_{p_{z}}=t_{p_{z}}(\phi_{01}+\phi_{11}+\phi_{10}+\text{h.c.})
\end{equation}
The coupling involving the $(\tau, p_{\pm})$ orbitals is more intricate due to the orbital structure's two-dimensional nature. This higher dimensionality introduces inter-orbitals couplings.

We define 
\begin{equation}
  C_{p_{\pm}p_{\pm}} = t_{p_{\pm}p_{\pm}}^{+}(\phi_{01}+\phi_{\bar{1}\bar{1}}\omega+\phi_{10}\omega^{*})
  +t_{p_{\pm}p_{\pm}}^{-}(\phi_{0\bar{1}}+\phi_{11}\omega+\phi_{\bar{1}0}\omega^{*}), \qquad \omega=e^{i\frac{2\pi}{3}}.
\end{equation}
We then have 
\begin{equation}
\hat{H}_{p_\pm}=t_{p_\pm}(\phi_{01}+\phi_{11}+\phi_{10}+h.c.)\begin{pmatrix} 1 & 0 \\ 0 & 1 \\ \end{pmatrix}+\begin{pmatrix} 0 & C_{p_\pm p_\pm}^\dagger \\ C_{p_\pm p_\pm} & 0 \\ \end{pmatrix}
\end{equation}
For the Kagome lattice we consider both a NN coupling $t_{\kappa}$ and a NNN coupling $t_{\kappa}'$, otherwise there would be an additional
flat band~\cite{li2018realization}.
\begin{equation}
\hat{H}_{\kappa}=t_{\kappa}\begin{pmatrix}0 & \phi_{\bar{1}0} & 1\\
1 & 0 & \phi_{0\bar{1}}\\
\phi_{11} & 1 & 0
\end{pmatrix}+t_{\kappa}'\begin{pmatrix}0 & \phi_{\bar{1}\bar{1}} & \phi_{\bar{1}0}\\
\phi_{0\bar{1}} & 0 & \phi_{10}\\
\phi_{01} & \phi_{11} & 0
\end{pmatrix}+\text{h.c}
\end{equation}
Finally we introduce couplings between the $(\tau,p_{z})$, $(\tau,p_{\pm})$
orbitals and the $(\tau,p_{\pm})$, $(\kappa,s)$ orbitals,
\begin{equation}
	\begin{split}
  \hat{C}_{p_{\pm}p_{z}} = & it_{p_{\pm}p_{z}}^{+}\begin{pmatrix}\phi_{01}+\phi_{\bar{1}\bar{1}}\omega+\phi_{10}\omega^{*}\\
-(\phi_{0\bar{1}}+\phi_{11}\omega^{*}+\phi_{\bar{1}0}\omega)
\end{pmatrix}
-it_{p_{\pm}p_{z}}^{-}\begin{pmatrix}\phi_{0\bar{1}}+\phi_{11}\omega+\phi_{\bar{1}0}\omega^{*}\\
-(\phi_{01}+\phi_{\bar{1}\bar{1}}\omega^{*}+\phi_{10}\omega)
\end{pmatrix}\\
 \hat{C}_{\kappa p_{\pm}} = & t_{\kappa p_{\pm}}^{+}\begin{pmatrix}\phi_{\bar{1}0} & \phi_{\bar{1}\bar{1}}\\
\phi_{\bar{1}\bar{1}}\omega^{*} & \omega\\
\omega &\phi_{\bar{1}0}\omega^{*} \\
\end{pmatrix}-t_{\kappa p_{\pm}}^{-}\begin{pmatrix}\phi_{\bar{1}\bar{1}} & \phi_{\bar{1}0}\\
\omega^{*} & \phi_{\bar{1}\bar{1}}\omega\\
\phi_{\bar{1}0}\omega & \omega^{*}
\end{pmatrix}
\end{split}
\end{equation}
\begin{table}[h]
\begin{tabular}{|c|c|c|}
\hline 
Parameter  & Meaning  & Ratio to $t_{\kappa}$\tabularnewline
\hline 
$\delta_{p_{z}}\equiv\mu_{p_{z}}+6t_{p_{z}}$  & $(\tau,p_{z})$ chemical potential  & -0.2593 \tabularnewline
$\delta_{p_{\pm}}\equiv\mu_{p_{\pm}}-3t_{p_{\pm}}$  & $(\tau,p_{\pm})$ chemical potential  & -0.3628 \tabularnewline
$\delta_{\kappa}\equiv\mu_{\kappa}+4(t_{\kappa}+t_{\kappa}')$  & $(\kappa,s)$ chemical potential  & 0.20\tabularnewline
$t_{p_{z}}$  & $(\tau,p_{z})$ NN  & 0.17 \tabularnewline
$t_{p_{\pm}}$  & $(\tau,p_{\pm})$ intra-orbital NN  & -0.03 \tabularnewline
$t_{p_{\pm}p_{\pm}}^{+}$  & $(\tau,p_{\pm})$ inter-orbital NN  & -0.065 \tabularnewline
$t_{p_{\pm}p_{\pm}}^{-}$  & $(\tau,p_{\pm})$ inter-orbital NN  & -0.055 \tabularnewline
$t_{\kappa}$  & $(\kappa,s)$ NN  & 1 \tabularnewline
$t_{\kappa}^{'}$  & $(\kappa,s)$ NNN  & 0.25 \tabularnewline
$t_{p_{\pm}p_{z}}^{+}$  & $(\tau,p_{\pm})-(\tau,p_{z})$ NN  & 0.095 \tabularnewline
$t_{p_{\pm}p_{z}}^{-}$  & $(\tau,p_{\pm})-(\tau,p_{z})$ NN  & 0.085 \tabularnewline
$t_{\kappa p_{\pm}}^{+}$  & $(\kappa,s)-(\tau,p_{\pm})$ NN  & 0.6 \tabularnewline
$t_{\kappa p_{\pm}}^{+}$  & $(\kappa,s)-(\tau,p_{\pm})$ NN  & 0.2 \tabularnewline
\hline 
\end{tabular}
\caption{Hopping parameters in the six-band model. We abbreviate nearest-neighbour with "NN" and next-to-nearest-neighbour with "NNN". We set the dominant energy scale to be
$t_{\kappa}=27$ meV, and measure all the other terms relative to
that. We use the parameters of Ref~\onlinecite{alvarado2023intrinsic}.}
\label{tab:parameter}
\end{table}

In what follows, for the sake of simplicity, we adopt the following convention.
Let $G(E,\mathbf{k})$ be any Green function in momentum space. Then we define $\mathbf{G}(E)\equiv G(E,\mathbf{k}) $ and
\begin{equation}
	G(E,\mathbf{R=0})=\int_{BZ} \frac{d^2\mathbf{k}}{S_{BZ}}G(E,\mathbf{k})
\end{equation}
We denote the local Green function $G(E,\mathbf{R=0})$ as $G(E)$ for simplicity.

\subsection{Low-energy Bogoliubov-de-Gennes Hamiltonian}\label{app:Low}
To simply understand some features of the subgap states, let's consider a 2 orbital model describing a $p_z$ and a $p_+$ orbital. We use the basis $(p_z)_{\uparrow}, (p_z)_{\downarrow},(p_+)_\uparrow, (p_+)_\downarrow$. Then the BdG Hamiltonian can be written as:
\begin{equation}
\hat{H}_{\rm BdG}=
\begin{pmatrix}
	\hat{H}_0 & \hat{\Delta} \\
	\hat{\Delta}^\dagger & -\hat{H}_0 \\
\end{pmatrix},
\quad
	\hat{H}_0(\mathbf{k})=\begin{pmatrix}
		\epsilon_{p_z}(\mathbf{k}) & \epsilon_{p_zp_+}(\mathbf{k}) \\
		\epsilon_{p_+p_z}(\mathbf{k}) & \epsilon_{p_+}(\mathbf{k}) \\
	\end{pmatrix}\otimes \mathds{1}_2, 
\quad
	\hat{\Delta}(\mathbf{k})=\begin{pmatrix}
	0 & 0 \\
	0 & \Delta(\mathbf{k}) \\
\end{pmatrix}\otimes\sigma_z
\end{equation}
For the simplicity of notation we drop the spin labels, and we define $\hat{H}^{\text{orb}}_0 \equiv \begin{pmatrix}
	\epsilon_{p_z}(\mathbf{k}) & \epsilon_{p_zp_+}(\mathbf{k}) \\
	\epsilon_{p_+p_z}(\mathbf{k}) & \epsilon_{p_+}(\mathbf{k}) \\
\end{pmatrix}$ and similarly $\hat{\Delta}^\text{orb}\equiv\begin{pmatrix}
	0 & 0 \\
	0 & \Delta(\mathbf{k}) \\
\end{pmatrix}$. Following the same steps as in the previous section we obtain for $\hat{\mathbf{G}}_e(\mathbf{k})$
\begin{equation}
	\begin{split}
		\hat{\mathbf{G}}_e(\mathbf{k}) & =\left[i\delta\mathds{1}_4-\hat{H}_0-\hat{\Delta} (i\delta\mathds{1}_4+\hat{H}_0)^{-1} \hat{\Delta}^{\dagger}\right]^{-1}(\mathbf{k})\\
		&  =\left[(i\delta\mathds{1}_2-\hat{H}_0^\text{orb})\otimes\mathds{1}_2-\hat{\Delta}^\text{orb}\otimes\sigma_z ((i\delta\mathds{1}_2+\hat{H}_0^\text{orb})\otimes\mathds{1}_2)^{-1} (\hat{\Delta}^\text{orb})^{\dagger}\otimes\sigma_z\right]^{-1}(\mathbf{k})\\
		&=\left[(i\delta\mathds{1}_2-\hat{H}_0^\text{orb})-\hat{\Delta}^\text{orb} (i\delta\mathds{1}_2+\hat{H}_0^\text{orb})^{-1} (\hat{\Delta}^\text{orb})^{\dagger}\right]^{-1}\otimes\mathds{1}_2(\mathbf{k})
	\end{split}
\end{equation}
For simplicity, in the following formulas we consider a quasiparticle inverse lifetime $\delta=0$, and we can write:

\begin{equation}\label{eq:Electronic_Green}
	\begin{split}
		\hat{\mathbf {G}}_e(\mathbf{k})= & \begin{pmatrix}
		\mathbf{G}_{p_z}(\mathbf{k}) & \mathbf{G}_{p_zp_+}(\mathbf{k}) \\
		\mathbf{G}_{p_zp_+}^\star(\mathbf{k}) & \mathbf{G}_{p_+}(\mathbf{k})
	\end{pmatrix}\otimes\mathds{1}_2 
	\\
	\hat{\mathbf{G}}_e(\mathbf{k})= & \begin{pmatrix}
		\frac{-\epsilon_{p_z}(\epsilon_{p_+}^2+|\Delta|^2)+\epsilon_{p_+}|\epsilon_{p_zp_+}|^2}{|\Delta|^2\epsilon_{p_z}^2+(\epsilon_{p_+}\epsilon_{p_z}-|\epsilon_{p_zp_+}|^2)^2} & \frac{\epsilon_{p_zp_+}}{-|\epsilon_{p_zp_+}|^2+\epsilon_{p_z}\left(\epsilon_{p_+}+\frac{|\Delta|^2\epsilon_{p_z}}{\epsilon_{p_+}\epsilon_{p_z}-|\epsilon_{p_zp_+}|^2}\right)}  \\
		\frac{\epsilon_{p_zp_+}^\star}{-|\epsilon_{p_zp_+}|^2+\epsilon_{p_z}\left(\epsilon_{p_+}+\frac{|\Delta|^2\epsilon_{p_z}}{\epsilon_{p_+}\epsilon_{p_z}-|\epsilon_{p_zp_+}|^2}\right)} & \frac{\epsilon_{p_z}(-\epsilon_{p_+}\epsilon_{p_z}+|\epsilon_{p_zp_+}|^2)}{|\Delta|^2\epsilon_{p_z}^2+(\epsilon_{p_+}\epsilon_{p_z}-|\epsilon_{p_zp_+}|^2)^2} 
	\end{pmatrix}(\mathbf{k})\otimes\mathds{1}_2
   \end{split}
\end{equation}
%
We show in Appendix~\ref{sec:FS_topology} that to accurately predict the polarization of bound states, it is crucial to consider the sign of the integrated Green functions. 
Because $p_z$ and $p_+$ belong to different irreducible representations of the model, $G_{p_zp+} \ll G_{p_z}$. Therefore, the spin polarization is entirely determined by the sign of $G_{p_z}$. 


Deep inside or outside the Fermi surface we can neglect the $\Delta$ term in the numerator of $G_{p_z}$, and the expression simplifies to:
\begin{equation}
	 \mathbf{G}_{p_z}(\mathbf{k})\sim \frac{-\epsilon_{p_z}\epsilon_{p_+}^2+\epsilon_{p_+}|\epsilon_{p_zp_+}|^2}{|\Delta|^2\epsilon_{p_z}^2+(\epsilon_{p_+}\epsilon_{p_z}-|\epsilon_{p_zp_+}|^2)^2}(\mathbf{k})
\end{equation}
It is useful to notice that at low doping, for $|\mu| < 1.3$, the $(p_+,p_-)$ orbitals dominate at the Fermi surface. This implies that in this regime, $|\epsilon_{p_z}(\mathbf{k})| \gg |\epsilon_{p_+}(\mathbf{k})|$. Additionally, we can assume that inter-orbital coupling is also small at low doping since the $(p_+,p_-)$ orbitals dominate the bands near the $K,K'$ points. With these approximations, we can simplify the analysis of the electronic Green functions for the low doping regime:
\begin{equation}\label{Eq:SimpleGpz}
	\mathbf{G}_{p_z}(\mathbf{k})\sim \frac{-\epsilon_{p_z}\epsilon_{p_+}^2}{|\Delta|^2\epsilon_{p_z}^2+(\epsilon_{p_+}\epsilon_{p_z}-|\epsilon_{p_zp_+}|^2)^2}(\mathbf{k}).
\end{equation}
The sign of $\mathbf{G}_{p_z}(\mathbf{k})$ is determined solely by the sign of $\epsilon_{p_z}(\mathbf{k})$. At low doping, the weight of the $p_z$ orbital is mainly supported by bands below the Fermi surface, which implies $-\epsilon_{p_z}(\mathbf{k}) \ge 0$ for almost all $\mathbf{k} \in BZ$. Consequently, $\mathbf{G}_{p_z}(\mathbf{k}) \ge 0$ for nearly all $\mathbf{k} \in BZ$, and therefore $G_{p_z} \ge 0$.
%
%
\section{Number of states\label{sec:Number_states}}
In our study we focus solely on the $(p_+,p_-,p_z)$ orbitals situated on the triangular lattice of the 6B1V model. This approximation is valid as the orbitals located on the Kagome lattice are not superconducting, and are far in energy from the $p$-orbitals, with little hybridization. Thus, they can only impact the number of bound states if the impurity acts directly on them, a scenario that we will not consider. In this section, we disregard the spin structure because as we do not introduce spin-orbit coupling it does not play a role in determining the dominant orbital element of the Green function for each pairing symmetry. We use the hat ($\hat{.}$) symbol to describe the matrix quantities in orbital space in order to differentiate them from the scalar quantities.
%

%


When an impurity acts on the $p_z$ orbital, it is probing a different physics compared to an impurity acting on either $p_+$ or $p_-$. This is because $p_z$ is a non-superconducting orbital. To distinguish between the superconducting and non-superconducting orbitals, we use a block decomposition with $p_z$ and $(p_+,p_-)$. In this block basis, the electron part of the impurity potential, $\hat{V}_e$, the electron/hole Green's function, $\hat{G}^{e/h}$, and the anomalous Green's function, $\hat{G}_\Delta$, respectively, take the form
\begin{equation}
	\hat{V}_e=\begin{pmatrix}
		V_{p_z} & \hat{0} \\
		\hat{0} & \hat{V}_{p_\pm} \\
	\end{pmatrix}, \qquad 
\begin{pmatrix}
\hat{G}^{e/h}=	G_{p_z}^{e/h} & \hat{G}_{p_z,p_\pm}^{e/h} \\
	\hat{G}_{p_\pm p_z}^{e/h} & \hat{G}_{p_\pm}^{e/h}\\
\end{pmatrix}, \qquad
\hat{G}_\Delta=\begin{pmatrix}
	G_{p_z,p_z}^\Delta & \hat{G}_{p_z,p_\pm}^{\Delta} \\
	\hat{G}_{p_\pm,p_z}^{\Delta} & \hat{G}_{p_\pm}^{\Delta}\\
\end{pmatrix}
\end{equation}
%
with $G^\Delta_{p_z,p_z}$ negligible for $d$- and $p$-wave, but crucial for ensuring the validity of Anderson theorem for $s$-wave pairing as shown in Appendix~\ref{sec:symmetry_swave}. 

A study of symmetries shows that the off-diagonal components of $\hat{G}^{e/h}$ vanish so that $\hat{G}^{e/h}$ simplifies to $\hat{G}^{e/h}={\rm blkdiag}\left( G_{p_z}^{e/h},  \hat{G}_{p_\pm}^{e/h} 
\right)$, where the notation "blkdiag" denotes the block diagonal operation.
Equation~\eqref{eq:Tmatrix_decomposition} shows that the electron/hole component of the T-matrix is given by 
 $\hat{T}_e=(\hat{A}_e+\hat{V}_e\hat{G}_\Delta \hat{A}_h^{-1}\hat{V}_e\hat{G}_{\Delta^\dagger})^{-1}\hat{V}_e =\hat{\mathcal{T}}_{\rm e}^{-1} \hat{V}_{\rm e}$
  with
   $\hat{A}_e=\mathds{1}-\hat{V}_e\hat{G}^e$ and $\hat{A}_h=\mathds{1}+\hat{V}_e\hat{G}^h$. 
   
The subgap states correspond to the pole of $\hat{T}_{\rm e}$, and come from the zeroes of $\hat{\mathcal{T}}_{\rm e}$. In all generality, the block components of this matrix have the following expression:

\begin{equation}
	\begin{split}
\left[\hat{\mathcal{T}}_{\rm e}\right]_{p_z,p_z} & = 1-V_{p_z}G_{p_z}^e+V_{p_z}\hat{G}_{p_z,p_\pm}^\Delta(\mathds{1}_2+\hat{V}_{p_\pm}\hat{G}_{p_\pm}^h)^{-1}\hat{V}_{p_\pm}\hat{G}_{p_\pm,p_z}^{\Delta^\dagger}\\
\left[\hat{\mathcal{T}}_{\rm e}\right]_{p_\pm,p_z} &= V_{p_z}\hat{G}_{p_z,p_\pm}^\Delta(\mathds{1}_2+\hat{V}_{p_\pm}\hat{G}_{p_\pm}^h)^{-1}\hat{V}_{p_\pm}\hat{G}_{p_\pm}^{\Delta^\dagger}\\
\left[\hat{\mathcal{T}}_{\rm e}\right]_{p_z,p_\pm} & = \hat{V}_{p_\pm}\hat{G}_{p_\pm}^\Delta(\mathds{1}_2+\hat{V}_{p_\pm}\hat{G}_{p_\pm}^h)^{-1}\hat{V}_{p_z}\hat{G}_{p_\pm,p_z}^{\Delta^\dagger}\\
\left[\hat{\mathcal{T}}_{\rm e}\right]_{p_\pm,p_\pm} & =  \mathds{1}_2-\hat{V}_{p_\pm}\hat{G}_{p_\pm}^e+\hat{V}_{p_\pm}\hat{G}_{p_\pm,p_z}^{\Delta}(1+V_{p_z}G_{p_z}^h)^{-1}V_{p_z}\hat{G}_{p_z,p_\pm}^{\Delta^\dagger}+\hat{V}_{p_\pm}\hat{G}_{p_\pm}^\Delta(1+\hat{V}_{p_\pm}\hat{G}_{p_\pm}^h)^{-1}\hat{V}_{p_\pm}\hat{G}_{p_\pm}^{\Delta^\dagger}
\label{eq:T_matrix_general}
	\end{split}
\end{equation}

%
%
To understand the results presented in Section~\ref{sec:LDOS} and Section~\ref{sec:SPLDOS} we will now consider some simplified cases that capture the physics leading to the apparition of subgap states.


\subsection{$p_z$ orbitals\label{sec:pz_orbital}}
\subsubsection{Superconducting state}
Let us consider the scenario where the impurity potential affects only the $p_z$ orbital, which can be represented as $\hat{V}_e=\begin{pmatrix}
V_{p_z} & \hat{0}_{1,2} \\
\hat{0}_{2,1} & \hat{0}_2 \\
\end{pmatrix}$. The poles of the $T$-matrix are the zeroes of $\hat{\mathcal{T}}_{\rm e}$ which now takes a simple form, from Eq.~\eqref{eq:T_matrix_general}:
\begin{equation}
\hat{\mathcal{T}}_{\rm e}=\begin{pmatrix}
    1-V_{p_z}G_{p_z}^e& \hat{0}_{1,2}\\ \hat{0}_{2,1} & \hat{1}_2 \\
\end{pmatrix} .
\label{eq:Tmatrix_pz}
\end{equation}

\begin{figure}[h]
\begin{center}
\begin{tikzpicture}
\node at (-5,0) {
\includegraphics[width=9cm]{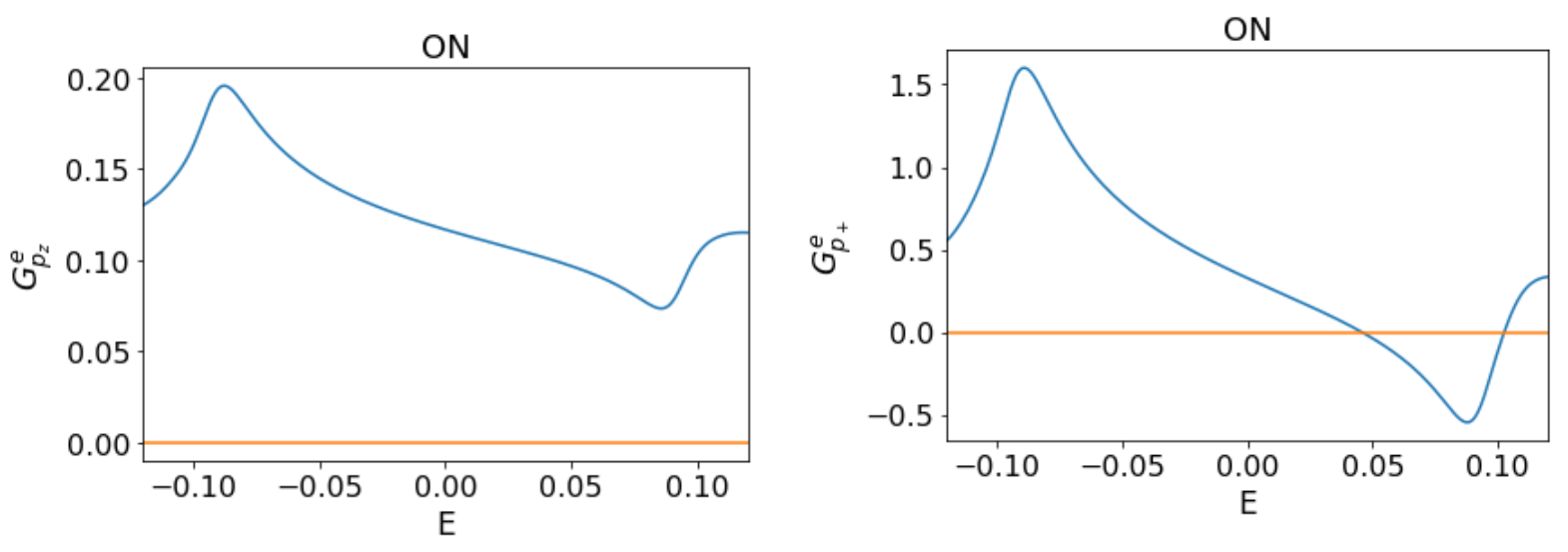}};
\node at (-8.3,1.4) {(a)};
\node at (-3.8,1.5) {(b)};
\end{tikzpicture}
\end{center}
\caption{$G^e$ components as a function of energy for an on-site pairing (blue line) and for (a) : a non SC $p_z$ orbital and (b) : a SC $p_+$ orbital. The crossings with the orange line indicate the zeroes of the $G^e$ components. We see that for (a) there is no crossing, whereas there is one for (b). We use $\Delta_0=0.1$ and $\mu=-0.4$.}
\label{fig:Ge_pz_pplus} 
\end{figure}

This expression reveals that the formation of subgap states is not affected by the hybridization between the $p_z$ and $p_\pm$ orbitals. As the $p_z$-orbital is not superconducting, $G_{p_z}^e(E)$ exhibits a different energy dependence, similar to that of the normal state compared to $\hat{G}_{p_\pm}^e(E)$, as shown in the main text. Specifically, $G_{p_z}^e(E)$ is always positive inside the gap because the $p_z$ orbital is non-superconducting and non-topological\cite{slager2015impurity}. On the contrary $\hat{G}_{p_\pm}^e(E)$ is positive at one gap edge and negative at the other, crossing zero as shown in Fig.~\ref{fig:Ge_pz_pplus} for an onsite $s$-wave superconducting state. This observation explains why only two states are seen for each fully gapped order parameter when the impurity exclusively affects the $p_z$ orbital. Since $G_{p_z}^e(E)$ is bounded between two positive values called $G_{min}$ and $G_{max}$ for every pairing symmetry within the gap, two particle-hole symmetric subgap will exist only for a finite range of the impurity potential. This stands in sharp contrast to the case where the impurity affects the superconducting orbitals $p_\pm$, as we will shown in what follows, for which the Green function zeroes enable the formation of subgap states for arbitrarily high impurity potentials. The significance of the zeroes of the integrated Green function was first emphasized in Ref.~\cite{slager2015impurity} to differentiate between trivial and topological insulators even though it has some limitations in more general situations~\cite{diop2020impurity}.

It is important to note that only for $s$-wave pairing, $\hat{G}^\Delta_{p_z,p_z}\ne 0$ as shown in Appendix~\ref{sec:symmetry_swave}. While this term may be much smaller than other components of the local anomalous Green function when the impurity affects the $p_+$ and $p_-$ orbitals, it is still crucial in ensuring the validity of Anderson's theorem when the impurity acts only on the $p_z$ orbital. In contrast, for $p$- and $d$-wave pairing, this term is zero due to symmetry considerations. In the case of $s$-wave pairing, Eq.~\eqref{eq:Tmatrix_pz} contains an additional term that must be taken into account
\begin{equation}
\hat{\mathcal{T}}_{\rm e}=
\begin{pmatrix}
    1-V_{p_z}G_{p_z}^e+V_{p_z}^2G_{p_z,p_z}^\Delta G_{p_z,p_z}^{\Delta^\dagger}(1+V_{p_z}G_{p_z}^h)^{-1}& \hat{0}_{1,2}\\ \hat{0}_{2,1} & \hat{1}_2 
\end{pmatrix} .
\end{equation}

\subsubsection{Normal state\label{sec:Normal_state_pz}}

We can understand the results shown in Fig.~\ref{fig:LDOS_Normal} using the same type of analysis as for the superconducting case. We now no-longer have particle-hole symmetry, and $\hat{\mathcal{T}_e}$ reduces to 
\begin{equation}
\mathcal{T}_e=1-V_{p_z}G_{p_z}^e
\end{equation}
To fully understand the spin polarization of the subgap states in the presence of a magnetic impurity, it is essential to explicitly consider the spin dependence. Let's consider a scenario where the impurity spin is aligned in the z-direction. In this case $V_{p_z}=\begin{pmatrix}1 & 0 \\ 0 & -1 \\\end{pmatrix}$ and because $G_{p_z}^e$ is spin degenerated,  $G_{p_z}^e=\begin{pmatrix} G_{p_z} &0 \\ 0 & G_{p_z} \\\end{pmatrix}$. We then have 
\begin{equation}
\mathcal{T}_e=\begin{pmatrix}
1-VG_{p_z} & 0 \\
0 & 1+VG_{p_z} \\
\end{pmatrix}.
\end{equation}
As illustrated in Fig.~\ref{fig:Ge_pz_pplus}, the quantity $G_{p_z}$ only takes positive values inside the gap, implying that a zero of $\mathcal{T}_e$ can only occur for the upper-spin component $1-VG{p_z}$. This observation provides an explanation for the equal contributions of both up and down states to the intensity of subgap states in the presence of a scalar impurity. Conversely, in the case of a magnetic impurity, only states with the same spin polarisation as the impurity can be subgap states.
\subsection{$(p_+,p_-)$ orbitals\label{sec:ppm_orbitals}}

Now let's consider the scenario where the impurity potential affects only the $(p_+,p_-)$ orbitals such that in the block basis $\hat{V}_e=\begin{pmatrix}
    0 & \mathbf{0}_{1,2} \\
    \mathbf{0}_{2,1} & \hat{V}_{p_\pm} \\
\end{pmatrix}$. For the rest of this section, we will restrict the block basis to the two orbitals $(p_+,p_-)$. 
We can express the electronic part of the T-matrix as follows:
\begin{equation}
\hat{T}_e=
    (\hat{A}_e+\hat{V}_{p_\pm}\hat{G}_\Delta A_h^{-1}\hat{V}_{p_\pm}\hat{G}_{\Delta^\dagger} )^{-1}.\hat{V}_{p_\pm}
\label{eq:Tmat_2orb}
\end{equation}
%
 with $\hat{A}_e=\mathds{1}_2-\hat{V}_{p_\pm}\hat{G}_{p_\pm}^e$, $\hat{A}_h=\mathds{1}_2+\hat{V}_{p_\pm}\hat{G}_{p_\pm}^h$, $\hat{G}_{p_\pm}^{e/h}=\begin{pmatrix}
    G_{p_+}^{e/h} & G_{p_+p_-}^{e/h} \\
    G_{p_-p_+}^{e/h} & G_{p_-}^{e/h} \\
\end{pmatrix}$ and $\hat{G}_{\Delta}=\begin{pmatrix}
    G_{p_+}^\Delta & G_{p_+p_-}^\Delta \\
    G_{p_-p_+}^{\Delta}& G_{p_-}^\Delta \\
\end{pmatrix}$. A study of symmetries again shows that the off-diagonal components of $\hat{G}^{e/h}$  vanish, so that $\hat{G}^{e/h}={\rm diag}\left( G_{p_+}^{e/h}, G_{p_-}^{e/h}\right)$
 
 We choose $\hat{V}_{p_\pm}$ diagonal in orbital space $\hat{V}_{p_\pm}={\rm diag}\left( V_1, V_2\right)$ and in general $V_1 \ne V_2$ so we also treat the case where the impurity potential is acting only on one of the two orbitals (by putting $V_1=0$ and $V_2\ne 0$ for example).\\
The analog of Eq.~\eqref{eq:T_matrix_general} is

\begin{equation}
	\begin{split}
\left[\hat{\mathcal{T}}_{\rm e}\right]_{p_+,p_+} & = 1-V_1G_{p_+}^e+V_1^2G_{p_+}^\Delta(1-V_1G_{p_+}^h)^{-1}G_{p_+}^{\Delta^\dagger}+V_1V_2G_{p_\pm}^\Delta(1+V_2G_{p_-}^h)^{-1}(G_{p_\pm}^{\Delta^\dagger})^\dagger \\
\left[\hat{\mathcal{T}}_{\rm e}\right]_{p_+,p_-} & = V_1V_2G_{p_+}^\Delta(1+V_1G_{p_+}^h)^{-1}G_{p_\pm}^{\Delta^\dagger}+V_1V_2G_{p_\pm}^\Delta(1+V_2G_{p_-}^h)^{-1}G_{p_-}^{\Delta^\dagger} \\
\left[\hat{\mathcal{T}}_{\rm e}\right]_{p_-,p_+} & = V_1V_2(G_{p_\pm}^{\Delta})^\dagger(1+V_1G_{p_+}^h)^{-1}G_{p_+}^{\Delta^\dagger}+V_1V_2G_{p_-}^\Delta(1+V_2G_{p_-}^h)^{-1}(G_{p_\pm}^{\Delta^\dagger})^\dagger \\
\left[\hat{\mathcal{T}}_{\rm e}\right]_{p_-,p_-} &= 1-V_2 G_{p_-}^e+V_2^2G_{p_-}^\Delta(1+V_2G_{p_-}^h)^{-1}G_{p_-}^{\Delta^\dagger}  +V_1V_2(G_{p_\pm}^\Delta)^\dagger(1+V_1G_{p_+}^h)^{-1}G_{p_\pm}^{\Delta^\dagger}
\label{eq:T_matrix_ppm}
     \end{split}
\end{equation}

To simplify the discussion, let us consider two limiting cases where the anomalous Green function is dominated either by the intra-orbital (diagonal : $G_{p_+}^\Delta$ and $G_{p_-}^\Delta$) terms or inter-orbital (off-diagonal $G_{p_+,p_-}^\Delta$ and $G_{p_-,p_+}^\Delta$) terms.\\

\subsubsection{Intra-orbital anomalous effect\label{sec:ppm_intra_orbital}}

When intra-orbital terms dominate the anomalous Green function, it means that $G^{\Delta}_{p_+}$ and $G^{\Delta}_{p_-}$ are much larger than $G^{\Delta}_{p_+p_-}$ and $G^{\Delta}_{p_-p_+}$. In this case, we can assume that the inter-orbital terms can be neglected, which simplifies the analysis of pairing symmetries. We show in Appendix \ref{sec:symmetry_swave} that all $s$-wave pairings satisfy these conditions. The local anomalous Green function then simplifies to $\hat{G}_\Delta={\rm blkdiag}\left(G_{p_+}^\Delta,G_{p_-}^\Delta \right)$
 and Eq.~\eqref{eq:T_matrix_ppm} simplifies to
\begin{equation}
	\begin{split}
		\hat{\mathcal{T}}_{\rm e} & = {\rm blkdiag}\left(\left[\hat{\mathcal{T}}_{\rm e}\right]_{p_+,p_+}, \left[\hat{\mathcal{T}}_{\rm e}\right]_{p_-,p_-}\right) \\
	\left[\hat{\mathcal{T}}_{\rm e}\right]_{p_+,p_+} & = 
		1-V_1G_{p_+}^e+V_1^2G_{p_+}^\Delta(1+V_1G_{p_+}^h)^{-1}G_{p_+}^{\Delta^\dagger} \\
		\left[\hat{\mathcal{T}}_{\rm e}\right]_{p_-,p_-} & = 1-V_2 G_{p_-}^e+V_2^2G_{p_-}^\Delta(1+V_2G_{p_-}^h)^{-1}G_{p_-}^{\Delta^\dagger}
\end{split}
\end{equation}
The matrix is diagonal in the $(p_+,p_-)$ basis, and it is clear that the diagonal components are equal for $V_1=V_2$, when $p_+$ and $p_-$ are degenerate. Thus, we can conclude that the spectrum will not change if we put the impurity on one of the two orbitals ($V_1\ne 0 $ and $V_2=0$) or on both orbitals ($V_1=V_2 \ne 0$). This observation is in agreement with the numerical results.


\subsubsection{Inter-orbital anomalous effect: type 1\label{sec:ppm_inter_orbital1}}

In the regime where inter-orbital terms dominate, the anomalous Green function components $G^{\Delta}_{p_+p_-}$ and $G^{\Delta}_{p_-p_+}$ are much larger than $G^{\Delta}_{p_+}$ and $G^{\Delta}_{p_-}$. To simplify the analysis, let us consider the extreme case where $G^{\Delta}_{p_+}=G^{\Delta}_{p_-}=0$. Pairing symmetries that belong to this class are called type 1. In Appendix \ref{sec:symmetry_pwave} and \ref{sec:symmetry_dwave}, we demonstrate that orbitals $p_x$, $p_y$, $d_{xy}$, and $d_{x^2-y^2}$ belong to this class. The local anomalous Green function can be simplified as : $\hat{G}_\Delta=\begin{pmatrix}
    0 & G_{p_+,p_-}^\Delta \\
    G_{p_-,p_+}^\Delta & 0 \\
\end{pmatrix}$ and Eq.~(\ref{eq:T_matrix_ppm}) simplifies to

\begin{equation}
	\begin{split}
		\hat{\mathcal{T}}_{\rm e} & = {\rm diag}\left(\left[\hat{\mathcal{T}}_{\rm e}\right]_{p_+,p_+}, \left[\hat{\mathcal{T}}_{\rm e}\right]_{p_-,p_-}\right) \\
		\left[\hat{\mathcal{T}}_{\rm e}\right]_{p_+,p_+} & = 1-V_1G_{p_+}^e+V_1V_2G_{p_\pm}^\Delta(1+V_2G_{p_-}^h)^{-1}(G_{p_\pm}^{\Delta^\dagger})^\dagger \\
		\left[\hat{\mathcal{T}}_{\rm e}\right]_{p_-,p_-} & = 1-V_2 G_{p_-}^e+V_1V_2(G_{p_\pm}^\Delta)^\dagger(1+V_1G_{p_+}^h)^{-1}G_{p_\pm}^{\Delta^\dagger}
	\end{split}
\end{equation}


The matrix remains diagonal in the $(p_+,p_-)$ basis, and it is evident that the diagonal components are equivalent for $V_1=V_2$ when $p_+$ and $p_-$ are degenerate as seen when the intra-orbital terms dominate. However, with the impurity potential acting on both orbitals, the diagonal terms of the T-matrix depend on both $V_1$ and $V_2$, and thus, depending on the details of the model and specific parameters, the subgap bound states can be modified or even disappear. This effect is expected to be more pronounced for fully gapped pairings, as gapless pairings can be affected by quasiparticle damping, which can make this effect difficult to detect.

\subsubsection{Inter-orbital anomalous effect: type 2\label{sec:inter_orbital2}}

To understand the results for all pairings, we need to consider a particular case that applies to both $p+ip'$- and $d+id'$-wave when inter-orbital effects dominate.  In this case the local anomalous Green function only has one non-zero element $\hat{G}_\Delta=\begin{pmatrix}
    0 & G_{p_+p_-}^\Delta \\
    0 & 0
\end{pmatrix}$. Pairing symmetries that belong to this class are called type 2. Then Eq.~(\ref{eq:T_matrix_ppm}) simplifies to 

\begin{equation}
\hat{A}_e+\hat{V}_{p_\pm}\hat{G}_\Delta \hat{A}_h^{-1}\hat{V}_{p_\pm}\hat{G}_{\Delta^\dagger}=\begin{pmatrix}
    1-V_1G_{p_+}^e+V_1V_2G_{p_+p_-}^\Delta(1+V_2G_{p_-}^h)^{-1}G_{p_-p_+}^{\Delta^\dagger}& 0 \\
0 & 1-V_2 G_{p_-}^e\\
\end{pmatrix},
\label{eq:type2inter}
\end{equation}
for which the diagonal components of $\hat{A}_e+\hat{V}_{p_\pm}\hat{G}_\Delta \hat{A}_h^{-1}\hat{V}_{p_\pm}\hat{G}_{\Delta^\dagger}$ are no longer degenerate. This results in the emergence of additional bound states when the impurity potential acts on both orbitals simultaneously ($V_1=V_2\ne 0$). In contrast, in the type 1 cases, these diagonal components remain equal even when both orbitals are affected by the impurity potential.

\vspace{1\baselineskip}

To generalize the conclusion to the more complicated case where both intra- and inter-orbital terms are non-zero, we need to consider the competition between these terms in determining the bound states. If the intra-orbital term dominates, the bound states will be mainly determined by the diagonal terms of $\hat{G}_\Delta$, and the same poles will be obtained regardless of whether the impurity acts on one or both orbitals. However, if the inter-orbital term dominates, the bound states will be mainly determined by the off-diagonal terms of $\hat{G}_\Delta$, and additional bound states can appear when the impurity potential is non-zero on both orbitals.

\vspace{1\baselineskip}

\section{Spin flip of spin triplet impurity states and Fermi surface topology\label{sec:FS_topology}}
\subsection{$(p_+,p_-)$ orbitals\label{sec:spin_flip_ppm}}
In what follows we will try to give some quantitative arguments to justify the impurity-states spin flip based on the Fermi-surface topology and using a simple 1-band model. Let's consider an $\eta=x$ superconducting triplet pairing. We can write the superconducting order parameter as 
\begin{equation}
\hat{\Delta}(\mathbf{k})=\Delta(\mathbf{k})\hat{\sigma}_z
\end{equation}
where we take the $z$ axis as the quantization axis and $\Delta(\mathbf{k})$ is the form factor of the considered symmetry. The argument does not depend on the specific form of $\Delta(\mathbf{k})$ so it extends to all triplet pairing. In the basis $\Psi_{\mathbf{k}}=(\psi_{\mathbf{k}\uparrow},\psi_{\mathbf{k}\downarrow},\psi^\dagger_{-\mathbf{k}\uparrow},\psi^\dagger_{-\mathbf{k}\downarrow})$ the BdG Hamiltonian can be written as: 
\begin{equation}
\hat{H}_{BdG}=
\begin{pmatrix}
\hat{H}_0 & \hat{\Delta} \\
\hat{\Delta}^\dagger & -\hat{H}_0 \\
\end{pmatrix}
\end{equation}
For simplicity let's suppose $\hat{H}_0$ is spin degenerate and diagonal 
\begin{equation}
\hat{H}_0(\mathbf{k})=\begin{pmatrix}
\epsilon(\mathbf{k}) & 0 \\
0 & \epsilon(\mathbf{k}) \\
\end{pmatrix}
\hspace{1\baselineskip}
\hat{\Delta}(\mathbf{k})=\begin{pmatrix}
\Delta(\mathbf{k}) & 0 \\
0 & -\Delta(\mathbf{k}) \\
\end{pmatrix}
\end{equation}
We can compute the Green's function at $E=0$, taking into account the quasiparticle lifetime $\delta$ with the replacement $E\rightarrow E+i\delta$. 
\begin{equation}\hat{\bf G}=
\begin{pmatrix}
\hat{\bf G}_e & \hat{\bf G}_\Delta \\
\hat{\bf G}_{\Delta^\dagger} & \hat{\bf G}_h \\
\end{pmatrix}    
\label{eq:GreenTotal}
\end{equation}
The electronic component of the Green function $\hat{\bf G}_e(\mathbf{k})$  can be written as:
\begin{equation}
\hat{\bf G}_e(\mathbf{k})=\left[i\delta\mathds{1}_2-\hat{H}_0-\hat{\Delta} (i\delta\mathds{1}_2+\hat{H}_0)^{-1} \hat{\Delta}^{\dagger}\right]^{-1}(\mathbf{k})
=-\frac{i\delta+\epsilon(\mathbf{k})}{\delta^2+|\Delta(\mathbf{k})|^2+\epsilon(\mathbf{k})^2}\mathds{1}_2
\label{eq:Electronic_Green}
\end{equation}

Deep inside or outside the FS, the expression simplifies to :
\begin{equation}
\hat{\bf G}_e(\mathbf{k})\sim -\big(\frac{1}{\epsilon(\mathbf{k})}+\frac{i\delta}{\epsilon(\mathbf{k})^2}\big)\mathds{1}_2.
\end{equation}
Thus the sign of the real part of $\mathbf{\hat{G}}_e(\mathbf{k})$ is given by the inverse of the sign of $\epsilon(\mathbf{k})$, which depends on $\mathbf{k}$ being inside or outside the FS, and the sign of the imaginary part is always negative. To understand the sign of the spin polarization $S_z$ given by Eq.~(\ref{eq:spin_components}) we compute 
\begin{equation}
\tilde{g}_{\sigma\sigma}(\mathbf{k})\sim \text{Im}\left[\hat{\bf G}(\mathbf{k})\hat{T}\hat{\bf G}(\mathbf{k})\right]_{e,\sigma\sigma},
\end{equation}
with $\hat{\mathbf{G}}$ given in Eq.~(\ref{eq:GreenTotal}).
We can simplify the analysis by studying only the non-anomalous terms of the T-matrix (which amounts to take $\hat{T}$ diagonal in particle-hole space) because a similar analysis can be performed for the off-diagonal terms of $\hat{T}$:
\begin{equation}
\tilde{g}_{\sigma_1\sigma_2}(E)= \text{Im}\big[\mathbf{\hat{G}}_e(E)\hat{T}_e(E)\mathbf{\hat{G}}_e(E)\\ 
\nonumber+\mathbf{\hat{G}}_{\Delta}(E)\hat{T}_h(E)\mathbf{\hat{G}}_{\Delta}^\dagger(E)\big]_{\sigma_1\sigma_2}.
\end{equation}
We call the first term $\mathbf{\hat{G}}_e\hat{T}_e\mathbf{\hat{G}}_e$ the electronic contribution and the second therm $\mathbf{\hat{G}}_\Delta \hat{T}_h\mathbf{\hat{G}}_\Delta^\dagger$ the hole contribution.
In the next section (Appendix \ref{sec:spin_polarization}), we demonstrate that the electronic and hole contributions are related through particle-hole symmetry. Therefore, it is sufficient to focus on studying the electronic contribution.
Because $\mathbf{\hat{G}}_e(\mathbf{k})$ has the following spin structure $\hat{\bf G}_e(\mathbf{k})\sim\mathds{1}_2$, we obtain
\begin{equation}
\tilde{g}_{\sigma\sigma'}(\mathbf{k})\sim \text{Im}\left[\left(\dfrac{1}{\epsilon(\mathbf{k})}+\dfrac{i\delta}{\epsilon(\mathbf{k})^2}\right)^2\begin{pmatrix}
    (\hat{T}_e)_{\uparrow\uparrow} & (\hat{T}_e)_{\uparrow\downarrow} \\
    (\hat{T}_e)_{\downarrow\uparrow} & (\hat{T}_e)_{\downarrow\downarrow} \\
\end{pmatrix}\right].
\end{equation}
This implies that the sign of $S_z$ depends on the relative value between $(\hat{T}_e)_{\uparrow\uparrow}$ and $(\hat{T}_e)_{\downarrow\downarrow}$. Recalling the T-matrix equation, we have 
\begin{equation}
\hat{T}_e=\big(\mathds{1}_2-\hat{V}_e\int \dfrac{d^2\mathbf{k}}{S_{BZ}}\hat{\bf G}_e(\mathbf{k})\big)^{-1} \hat{V}_e,
\end{equation}
where $\hat{V}_e$ is the z-impurity matrix, and can be written in the $(\uparrow,\downarrow)$ basis
\begin{equation}
 \hat{V}_e=\begin{pmatrix}
V_z & 0 \\
0 & -V_z \\
\end{pmatrix}.
\end{equation}
We choose $V_z \ge 0$. As $(\hat{\bf G}_{e})_{\uparrow\uparrow}=(\hat{\bf G}_{e})_{\downarrow\downarrow}$, we can write $\hat{\bf G}_{e{\uparrow\uparrow}}\equiv \bf G_e$.
There are then three possible situations to consider:

\noindent1. $\text{Re}\left[\mathlarger{\int}_{BZ} \dfrac{d^2\mathbf{k}}{S_{BZ}}{\bf G}_e(\mathbf{k})\right]=0$ : there can be no poles, so no impurity states in this trivial case. \\
2. $\text{Re}\left[\mathlarger{\int}_{BZ} \dfrac{d^2\mathbf{k}}{S_{BZ}}{\bf G}_e(\mathbf{k})\right]>0$ : there exists a critical value of $V_z$, such that $1-V_z\text{Re}\left[\mathlarger{\int}_{BZ} \dfrac{d^2\mathbf{k}}{S_{BZ}}{\bf G}_e(\mathbf{k})\right]=0$. In this case 
\begin{flalign}
\hat{T}_e=\begin{pmatrix}
    i(V_zA)^{-1}& 0 \\
    0 & 1/(iV_zA+1+V_z\text{Re}[\int_{BZ} {\bf G}_e(\mathbf{k})] \\
\end{pmatrix},
\end{flalign}
with $A=\text{Im}\left[\mathlarger{\int}_{BZ}\dfrac{d^2\mathbf{k}}{S_{BZ}}{\bf G}_e(\mathbf{k})\right]\ge 0$. In the limit of $\delta \rightarrow 0$, $A\rightarrow 0$ and $|(\hat{T}_e)_{\uparrow\uparrow}|\gg |(\hat{T}_e)_{\downarrow\downarrow}|$. In this limit
\begin{flalign}
\tilde{g}_{\sigma\sigma'}(\mathbf{k}) \sim \dfrac{1}{\epsilon(\mathbf{k})^2} \begin{pmatrix}
    \text{Im}[(\hat{T}_e)_{\uparrow\uparrow}] & 0 \\
    0 & 0 \\
\end{pmatrix},
\end{flalign} and we can conclude $S_z >0$ \\
3. $\text{Re}\left[\mathlarger{\int}_{BZ} \dfrac{d^2\mathbf{k}}{S_{BZ}}{\bf G}_e(\mathbf{k})\right]<0$ : we can follow the same reasoning to conclude $S_z<0$.


Although we use a simple one-band model for clarity, our findings can be extended to multi-band models such as the 6B1V model that take into account the relevant orbitals. In the 6B1V model, the Fermi surface centered around the $K$ point becomes centered around the $\Gamma$ point at $\mu \approx -0.63$, as shown in Fig. \ref{fig:FS_6B1V}. Consequently, $\text{Re}\left[\int_{BZ} d^2\mathbf{k} {\bf G}_e(\mathbf{k})\right]$ changes sign, and for every triplet pairing $\eta = x$, the change in the Fermi surface topology can cause a spin flip in the average polarization $S_z$ for a z-magnetic impurity. These results are consistent with those shown in Fig. \ref{fig:SPDOS_mu_pip}, indicating a spin flip as the Fermi surface centered around $K$ becomes centered around $\Gamma$. It is noteworthy that a local impurity effect can strongly depend on the global topology of the Fermi surface. This phenomenon can be generalized to any model with a similar change in the Fermi surface topology, and with superconducting triplet pairing.

It is also worth noting that the Fermi surface crossing or not the boundary of the first BZ is not relevant for the spin polarization, and therefore does not allow for the detection of a topological phase transition which occurs when the FS is touching the BZ boundary.


\subsection{$p_z$ orbital\label{sec:spin_flip_pz}}
For a $\eta=x$ triplet pairing, $S_z$ has a definite polarization, but contrary to the case where the impurity potential is on the $p_+/p_-$ orbitals, there is no spin flip as a function of the chemical potential $\mu$ as detailed in Sec.\ref{sec:FS_topology}. This can be epxlained by studying $\hat{G}_e$ in a 2-band model with $p_z$ and $p_+$ orbtials where only $p_+$ is a superconducting orbital.
For simplicity let's suppose $\hat{H}_0$ is spin degenerate and for generality let's allow a coupling between the $p_z$ and $p_+$ orbitals. In this section, the hat symbol ($\hat{.}$) denotes a matrix structure that can exist in both orbital and spin space.

Referring back to the T-matrix equation
\begin{equation}
\hat{T}_e=\big(\mathds{1}_4-\hat{V}_e\int \frac{d^2\mathbf{k}}{S_{BZ}}\hat{\mathbf{G}}_e(\mathbf{k})\big)^{-1} \hat{V}_e,
\end{equation}
$\hat{V}_e$ is the z-impurity matrix, and can be written in the $(p_z)_{\uparrow}, (p_z)_{\downarrow},(p_+)_\uparrow, (p_+)_\downarrow$ basis as
\begin{equation}
 \hat{V}_e=\begin{pmatrix}
V_{p_z} & 0 \\
0 & 0 \\
\end{pmatrix}\otimes\sigma_z
\end{equation}
We choose $V_{p_z} \ge 0$ and we get 
\begin{equation}
\hat{V}_e\hat{G}_e=\begin{pmatrix}
V_{p_z}G_{p_z} & V_{p_z}G_{p_zp_+} \\
0 & 0 \\
\end{pmatrix}\otimes\sigma_z.
\end{equation}
To understand why the spin polarization does not flip at low doping when the impurity potential acts only on the $p_z$ orbital, we consider the simplified $\mathbf{G}_{p_z}(\mathbf{k})$, Eq.~\eqref{Eq:SimpleGpz}.
In this low doping regime, $|\epsilon_{p_z}(\mathbf{k})| \gg |\epsilon_{p_+}(\mathbf{k})|$. Additionally, we can assume that inter-orbital coupling is also small at low doping since the $(p_+,p_-)$ orbitals dominate the bands near the $K,K'$ points.
The sign of $\mathbf{G}_{p_z}(\mathbf{k})$ is determined solely by the sign of $\epsilon_{p_z}(\mathbf{k})$. At low doping, the weight of the $p_z$ orbital is mainly supported by bands below the Fermi surface (FS), which implies $-\epsilon_{p_z}(\mathbf{k}) \ge 0$ for almost all $\mathbf{k} \in BZ$. Consequently, $\mathbf{G}_{p_z}(\mathbf{k}) \ge 0$ for nearly all $\mathbf{k} \in BZ$, and therefore $G_{p_z} \ge 0$.

\vspace{1\baselineskip}

In the previous section where the impurity potential acted on the $(p_+,p_-)$ orbitals, the spin polarization was strongly dependent on the chemical potential $\mu$ at low doping. However, when the impurity is only acting on the $p_z$ orbital, the constant sign of $G_{p_z}$ at low doping indicates that the spin polarization is independent of $\mu$. This behavior is observed numerically, where the spin polarization does not flip as the FS center shifts from the $K$ points to the $\Gamma$ point as shown in Fig.\ref{fig:SPDOS_mu_pip}.(c).

\vspace{1\baselineskip}

\section{Spin polarisation of subgap bound states\label{sec:spin_polarization}}
In this section we show how to predict the spin polarisation of opposite energy subgap bound states for every pairing symmetries.\\

We consider again a 1-band model where the orbital structure is trivial, having a dimension of 1. Therefore, the symbol $\hat{.}$ on the electronic, hole, and anomalous Green functions is used here to indicate the spin structure.
We can write the superconducting order parameter as 
\begin{equation}
\hat{\Delta}=\Delta(\mathbf{k})\hat{\sigma}
\label{eq:defSigma}
\end{equation}
where we take the $z$ axis as the spin quantization axis, $\Delta(\mathbf{k})$ the superconducting form factor of the considered symmetry and $\hat{\sigma}$ is an arbitrary Pauli matrix. The argument does not depend on the specific form of $\Delta(\mathbf{k})$ so we will omit the $\mathbf{k}$ dependence in the following. Applying the same step as in Appendix.\ref{sec:FS_topology},we get for the spin structure of the electronic, hole and anomalous Green function $\hat{\bf G}_e \sim \mathds{1}_2$, $\hat{\bf G}_h \sim \mathds{1}_2$, $\hat{\bf G}_\Delta \sim \hat{\sigma}$, $\hat{\bf G}_{\Delta^\dagger} \sim \hat{\sigma}^\dagger$ with $\sigma$ an arbitrary Pauli matrix as defined in Eq.\eqref{eq:defSigma}.

As before, the impurity potential is diagonal in the electron/hole basis: \begin{equation}\hat{V}=
\begin{pmatrix}
 \hat{V}_e & 0 \\ 
 0 & -\hat{V}_e \\
\end{pmatrix}.
\end{equation}
We will consider impurities with spin along x and z such that respectively $\hat{V_e}^x=\sigma_x$ and $\hat{V_e}^z= \sigma_z$.
Let us recall the T-matrix equation for the electronic and hole parts, while omitting the explicit dependence on energy:
\begin{align}
&\hat{T}_e=\bigg[\mathds{1}_2-\hat{V}_e\hat{G}_e+\hat{V}_e\hat{G}_\Delta(\mathds{1}_2+\hat{V}_e\hat{G}_h)^{-1}\hat{V}_e\hat{G}_{\Delta^\dagger}\bigg]^{-1} \hat{V}_e, &\\
&\hat{T}_h=-\bigg[\mathds{1}_2+\hat{V}_e\hat{G}_h+\hat{V}_e\hat{G}_{\Delta^\dagger}(\mathds{1}_2-\hat{V}_e\hat{G}_e)^{-1}\hat{V}_e\hat{G}_\Delta\bigg]^{-1} \hat{V}_e. &
\label{eq:Tmatrix_decomposition}
\end{align}
Subsequently, using the fact that $\hat{G}_h(E)=-\hat{G}_e(-E)^\dagger$ and $\hat{G}_\Delta(E)=[\hat{G}_{\Delta^\dagger}(-E)]^\dagger$, stemming from particle-hole symmetry, and that $\hat{V}_e^\dagger=\hat{V}_e$, we get 
\begin{flalign}
&\hat{T}_e(E)=-\hat{T}_h(-E)^\dagger \\
&\hat{T}_\Delta(E)=-[\hat{T}_{\Delta^\dagger}(-E)]^\dagger.
\label{eq:Tmatrix_ph}
\end{flalign}
So the key observation is that particle-hole symmetry implies that if $\hat{T}_e$ has a pole at $E$, then $\hat{T}_h$ will have a pole at $-E$. 

We recall that the spin polarization of the bound states $S_z$ and $S_x$ is given in terms of the T-matrix by $S_z(\mathbf{k},E) =\int_{BZ} d^2\mathbf{k}\big(\tilde{g}_{\uparrow\uparrow}(\mathbf{k},E)-\tilde{g}_{\downarrow\downarrow}(\mathbf{k},E)\big)$ and $S_x(\mathbf{k},E) = \int_{BZ} d^2\mathbf{k} \left(\tilde{g}_{\uparrow\downarrow}(\mathbf{k},E)+\tilde{g}_{\downarrow\uparrow}(\mathbf{k},E)\right)$ up to irrelevant prefactors where 
\begin{flalign}
&\tilde{g}_{\sigma_1\sigma_2}(E)\sim \text{Im}\left[\hat{\mathbf{G}}(E)\hat{T}(E)\hat{\mathbf{G}}(E)\right]_{e,\sigma_1\sigma_2},
\label{eq:gsigma}
\end{flalign} 
up to prefactors that are also irrelevant for the argument.


We can simplify this analysis by studying only the non-anomalous terms of the T-matrix (which amounts to taking $\hat{T}$ diagonal) because a similar analysis can be performed for the off-diagonal terms of $\hat{T}$:
\begin{flalign}
\tilde{g}_{\sigma_1\sigma_2}(E)= \text{Im}\big[\mathbf{\hat{G}}_e(E)\hat{T}_e(E)\mathbf{\hat{G}}_e(E)+\mathbf{\hat{G}}_{\Delta}(E)\hat{T}_h(E)\mathbf{\hat{G}}_{\Delta}^\dagger(E)\big]_{\sigma_1\sigma_2}.
\end{flalign}
 We first consider an impurity with spin along z and the potential
$\hat{V}_e=V \sigma_z$. The argument in Section~\ref{sec:FS_topology} shows that at a pole $E$, $\hat{T}_e(E)$ will have a definite polarisation $\sigma$.
Let's assume $\sigma=\uparrow$. Then $\hat{T}_e(E)= \begin{pmatrix}
    T_\uparrow & 0 \\
    0 & 0 \\
\end{pmatrix}$. \\
At the energy $E$, the  electronic part of $\tilde{g}$ dominates so extracting the spin dependence we get 
\begin{equation}
\mathbf{\hat{G}}_e(E) \hat{T}_e(E)\mathbf{\hat{G}}_e(E) =\begin{pmatrix}
T_\uparrow & 0 \\
0 & 0\\
\end{pmatrix}\big[\mathbf{G}_e(E)\mathbf{G}_{e}(E)\big].
\label{eq:ElectronicPolarisation}
\end{equation}
Particle-hole symmetry imposes that $\hat{T}_h$ has a pole at $-E$.
At energy $-E$, $\tilde{g}$ is dominates by the hole part $\mathbf{\hat{G}}_\Delta(-E)\hat{T}_h(-E)\mathbf{\hat{G}}_\Delta(-E)^\dagger$. Because $\mathbf{\hat{G}}_{\Delta}$ has the same spin dependence than $\mathbf{\hat{\Delta}}$, the spin singlet or triplet nature of the pairing will impose the polarization of the bound states at energy $-E$.
First consider the spin singlet case, $\hat{\Delta}\sim i\sigma_y$ for which we obtain:
\begin{flalign}
\mathbf{\hat{G}}_\Delta(-E)\hat{T}_h(-E)\mathbf{\hat{G}}_{\Delta^\dagger}(-E) =&\begin{pmatrix} 
0 & 0 \\
0 & T_\uparrow\\
\end{pmatrix}\big[\mathbf{G}_\Delta(-E)\mathbf{G}_{\Delta^\dagger}(-E)\big].
\label{eq:HolePolarisation}
\end{flalign}
The crucial point is that in both Eq.~(\ref{eq:ElectronicPolarisation}) and Eq.~(\ref{eq:HolePolarisation}), the scalar terms (respectively $\mathbf{G_e}(E)^2$ and $\mathbf{G}_\Delta(-E)\mathbf{G}_{\Delta^\dagger}(-E)$) have the same sign when integrated over the BZ,
\begin{flalign}
sign\left(\int d^2\mathbf{k}\mathbf{G}_e(E)^2 \right)= sign\left(\int d^2\mathbf{k}(\mathbf{G}_{\Delta}(-E)\mathbf{G}_{\Delta^\dagger}(-E))\right).
\label{eq:key_equation}
\end{flalign}
This means that what matters for determining the polarisation sign of the subgap bound states in the expression of $\tilde{g}$ is uniquely the matrix structure. For the state at $E$ the spin part is $\begin{pmatrix}
T_\uparrow & 0 \\
0 & 0 \\
\end{pmatrix}$, and for  the state at $-E$ it is $\begin{pmatrix}
0 & 0 \\
0 & \tilde{T}_\uparrow \\
\end{pmatrix}$,
where in principle $T_\uparrow$ and $\tilde{T}_\uparrow$ can be different but have the same sign.
Looking at the formula for $S_z$, this means the states at $E$ and $-E$ have opposite polarization.


For an $\eta=x$ spin triplet once more only the matrix form of $\tilde{g}$ will matter. In this case $\hat{\Delta}\sim \sigma_z$ which leads to 
\begin{flalign}
\mathbf{\hat{G}}_\Delta(-E)\hat{T}_h(-E)\mathbf{\hat{G}}_{\Delta^\dagger}(-E) =\begin{pmatrix}
  T_\uparrow & 0 \\  
  0 & 0\\
\end{pmatrix}\big[\mathbf{G}_{\Delta}(-E)\mathbf{G}_{\Delta^\dagger}(-E)\big]
\end{flalign}
So contrary to the spin singlet case, the states at $E$ and $-E$ will have the same polarization.


Now let the impurity spin be aligned along x so $\hat{V}_e=\sigma_x$. Because spin singlet SC are isotropic in spin space, let's consider only the spin triplet $\eta=x$ case. Again only the spin dependence of $\tilde{g}$ matters.\\ In this case with $\hat{V}_e =V\sigma_x$ and using Eq.~(\ref{eq:Tmatrix_decomposition}) we get
\begin{equation}
T_e(E)=\begin{pmatrix} a_1 & a_2 \\
a_3  & a_1 \\ 
\end{pmatrix},
\end{equation}
where $a_1$,$a_2$,$a_3$ are introduced to parameterize the spin components of $T_e$ with the condition that $(T_e)_{\uparrow\uparrow}=(T_e)_{\downarrow\downarrow}$ because the polarization has to be $0$ on the z axis.

The state at energy $E$ is dominated by the electronic contribution and we obtain $\hat{G}_{e}\hat{T}_e\hat{G}_{e} \sim A\begin{pmatrix}
    a_1  & a_2 \\
    a_3 & a_1 \\
\end{pmatrix}$ with $A$ an irrelevant prefactor that can be computed by integrating the product of electronic Green functions over the BZ. This leads to $S_x \sim \tilde{g}_{\uparrow\downarrow}+\tilde{g}_{\downarrow\uparrow} \sim A(a_2+a_3)$.


At $-E$, the hole parts dominates. For the spin triplet case $\eta=x$, $\hat{G}_{\Delta}\sim \hat{G}_{\Delta^\dagger} \sim \sigma_z$, thus we get $G_{\Delta}T_hG_{\Delta^\dagger} \sim \tilde{A}\begin{pmatrix}
    a_1 & -a_2 \\
    -a_3 & a_1\\
\end{pmatrix}$, where $\tilde{A}$ is another irrelevant prefactor. This leads to $S_x \sim -\tilde{A}(a_2+A_3)$. $A$ can be different than $\tilde{A}$ but Eq.~(\ref{eq:key_equation}) shows that $sign(A)=sign(\tilde{A})$. 
We can conclude that the states at $E$ and $-E$ have opposite polarization in this case.


\section{Symmetry diagnosis \label{sec:Symmetry_diagnosis}}
To investigate inter-orbital effects, we aim to determine when the off-diagonal components of the average electronic or anomalous Green function over the Brillouin Zone are non-zero. This corresponds to a sum of the form $\dfrac{1}{N}\sum_{\mathbf{k}}\mathbf{G}_{ij}(,E)=G_{ij}(\mathbf{R}=\mathbf{0},E)$, where $i$ and $j$ are orbital indices. The average can be interpreted as an on-site hopping between orbitals $i$ and $j$ \cite{ogata2013impurity}, and its magnitude depends on the symmetry of the two orbitals and the lattice. As the non trivial physics leading to inter-orbital effects only comes in our study from the orbitals located on the triangular lattice, we will ignore the orbitals located on the Kagome lattice in the following.

We follow the method of Ref.~\onlinecite{fischer2013gap} to understand the symmetry transformation of the BdG Hamiltonian.
The mean field Hamiltonian in the particle-hole basis  can be written as:
\begin{flalign}
\nonumber\hat{H}_{BdG}(\mathbf{k})=\begin{pmatrix}
\hat{H}_0(\mathbf{k}) & \hat{\Delta}(\mathbf{k}) \\
\hat{\Delta}^\dagger(\mathbf{k}) & -\hat{H}_0(\mathbf{k}) \\
\end{pmatrix}.
\end{flalign}
Because we consider only intra-orbital pairings we have 
\begin{flalign}
\hat{\Delta}(\mathbf{k})=\mathds{1}_2\psi(\mathbf{k}).
\end{flalign}
This implies that under a transformation of the point symmetry group $g$, $\hat{\Delta}(\mathbf{k})$ transforms as
\begin{flalign}
g[\hat{\Delta}(\mathbf{k})]=\sigma_0\psi(R_g\mathbf{k}),
\end{flalign}
with $R_g$ the corresponding rotation matrix in momentum space.
We can conclude that the transformation of $\hat{\Delta}(\mathbf{k})$ is directly given by the symmetry classification of the irreducible representation of the symmetry point group $D_6$ of the triangular lattice.

The magnetic point group of the single-valley model is $6^\prime2^\prime2$ and is generated by $C_{2z}\mathcal{T}$, $C_{3z}$ and a 2D rotation $C_{2x}$ which flips the $y$ coordinate. The diagonal part of $\hat{H}_0(\mathbf{k})$ transforms trivially in the $(p_z,p_+)$ space but the off-diagonal components transform non-trivially under the symmetry point group because the $(p_+,p_-)$ orbitals belongs to the $2D$ irreducible representation $E$.

It can be shown\cite{lessnich2021elementary} that the electronic Green's function transforms in the same way as the normal state Hamiltonian $H_0(\mathbf{k})$. Because only the diagonal components of $\hat{H}_0(\mathbf{k})$ transform trivially under symmetries of $6'2'2$, this means that the local electronic Green's function is diagonal in the $(p_+,p_-)$ basis:

\begin{equation}
\nonumber\dfrac{1}{N}\sum\limits_\mathbf{k}\hat{\mathbf{G}}_e(E)=
\begin{pmatrix}
G^e_{p_+}(\mathbf{R}=\mathbf{0}) & 0\\
 0 &G^e_{p_-}(\mathbf{R}=\mathbf{0})\\
\end{pmatrix}.
\end{equation}

\vspace{1\baselineskip}


Having understood the orbital structure of the local electronic Green's function $\hat{G}_e(\mathbf{R}=\mathbf{0})$, we need to apply the same analysis to the local anomalous Green's function $\hat{G}_\Delta(\mathbf{R}=\mathbf{0})$. We have the relationship:
\begin{flalign}
\hat{\mathbf{G}}_{\Delta}(E)=-\hat{\mathbf{G}}_e(E)\hat{\Delta}(\mathbf{k})(\hat{E}+\hat{H}_0(\mathbf{k}))^{-1}.
\end{flalign}
Let's define $\hat{\mathbf{G}}_e^0(E)\equiv (\hat{E}+\hat{H}_0(\mathbf{k}))^{-1}$ in the following for simplicity.
Let $i,j$ be two orbitals indices. Then we can write
\begin{flalign}
\hat{\mathbf{G}}_{\Delta}(E)_{ij}=-\sum\limits_{\alpha,\beta}\hat{\mathbf{G}}_e(E)_{i\alpha}\hat{\Delta}(\mathbf{k})_{\alpha\beta}\hat{\mathbf{G}}_e^0(E)_{\beta j} =-\sum\limits_{\alpha=p_\pm}\hat{\mathbf{G}}_e(E)_{i\alpha}\psi(\mathbf{k})\hat{\mathbf{G}}_e^0(E)_{\alpha j}.
\label{eq:symmetry_orbitals}
\end{flalign}
We use the fact that $\hat{\Delta}(\mathbf{k})=\sigma_0\psi(\mathbf{k}) \; (\sigma_0\vec{\psi}(\mathbf{k}))$ for singlet (triplet) pairing is diagonal in orbital space and non zero only for the $p_\pm$ low-energy orbitals.
The expression of $\mathbf{\hat{G}}_e$ and $\mathbf{\hat{G}}^0_e$ shows that the two quantities transform similarly under a point group transformation.

\subsubsection{$s$-wave pairing\label{sec:symmetry_swave}}
Both ON and $s_{ext}$ pairing belong to the $A_1$ irreducible representation of $D_6$ and are isotropic in $\mathbf{k}$ space which means that Eq.~(\ref{eq:symmetry_orbitals}) can now be written as, leaving the $E$ and $\mathbf{k}$ dependence implicit:
\begin{flalign}
(\hat{\mathbf{G}}_\Delta)_{ij}\sim -\sum\limits_{\alpha=p_\pm}(\hat{\mathbf{G}}_e)_{i\alpha}(\hat{\mathbf{G}}_e^0)_{\alpha j}.
\end{flalign}
This is because $\psi(\mathbf{k})$ transforms trivially under the spatial transformation.
For $i=j=p_\pm$ the integration over the BZ will give rise to a non-zero contribution because $(\hat{\mathbf{G}}_e)_{ii}$ and $(\hat{\mathbf{G}}_e^0)_{i i}$ are both invariant under the symmetries of $6'2'2$, so their product is also invariant.

We now look at the off-diagonal components. With no loss of generality let's assume $i=p_+$ and $j=p_-$. Then we have 
\begin{flalign}
\nonumber(\hat{\mathbf{G}}_\Delta)_{p_+p_-}\sim (\hat{\mathbf{G}}_e)_{p_+p_+}(\hat{\mathbf{G}}_e^0)_{p_+p_-}+(\hat{\mathbf{G}}_e)_{p_+p_-}(\hat{\mathbf{G}}_e^0)_{p_-p_-}.
\end{flalign}
Both terms are not invariant under the point group transformation so this means the off-diagonal components of the local anomalous Green function for $s$-wave pairing are negligible. Similar results hold for any $i,j \in \{p_+,p_-\}$ with $i \ne j$.

To understand the results for a scalar impurity acting on the $p_z$ orbital for $s$-wave pairing we need to understand why $(\hat{\mathbf{G}}_\Delta)_{p_z,p_z}$ is non vanishing. We have
\begin{flalign}
\nonumber(\hat{\mathbf{G}}_\Delta)_{p_zp_z}\sim (\hat{\mathbf{G}}_e)_{p_zp_+}(\hat{\mathbf{G}}_e^0)_{p_+p_z}+(\hat{\mathbf{G}}_e)_{p_zp_-}(\hat{\mathbf{G}}_e^0)_{p_-p_z}.
\end{flalign}
The invariance of both terms under the point group transformation implies that we cannot assume that $(\hat{\mathbf{G}}_\Delta)_{p_z,p_z}$ is zero for $s$-wave pairing based solely on symmetry considerations. Although $(\hat{\mathbf{G}}_\Delta)_{p_z,p_z}$ may be significantly smaller than other components of the local anomalous Green function, its non-vanishing nature is crucial when the impurity only affects the $p_z$ orbital where $(\mathbf{\hat{G}}_\Delta)_{p_{+/-},p_{+/-}}$ does not play any role as shown in Eq.~(\ref{eq:Tmatrix_pz}). The validity of Anderson's theorem relies precisely on $(\mathbf{\hat{G}}_\Delta)_{p_z,p_z}\ne 0$. For $p$- and $d$-wave pairing this term is zero by symmetry because they transform non-trivially under the point group transformation contrary to the $s$-wave pairing.

\subsubsection{$p$-wave pairing\label{sec:symmetry_pwave}}
The $p$-wave pairing belongs to the $E_1$ 2D irreducible representation of the $D_6$ point group symmetry. In contrast to the $s$-wave case, the transformation properties of $\hat{\psi}(\mathbf{k})$ are nontrivial under point group symmetries. For the case of $p+ip'$ pairing, we can express $\vec{\psi}(\mathbf{k})$ as $\vec{\psi}(\mathbf{k})\sim \psi_{p_+}$ to keep only the momentum dependence and ignore the spin singlet/triplet nature of the pairing. When $i=j$, say $i=p_+$, we have,
\begin{flalign}
(\hat{\mathbf{G}}_\Delta)_{p_+p_+}\sim  (\hat{\mathbf{G}}_e)_{p_+p_+}\psi_{p_+}(\hat{\mathbf{G}}_e^0)_{p_+p_+}+(\hat{\mathbf{G}}_e)_{p_+p_-}\psi_{p_+}(\hat{\mathbf{G}}_e^0)_{p_-p_+}.
\end{flalign}
Both terms transforms as $p_+$ and so are not invariant under the point group transformation, which means that, contrary to the $s$-wave case, the diagonal components of $\hat{G}_\Delta$ are suppressed for the $p$-wave symmetry.
We then consider $i\ne j$. We need to treat separately the 2 off-diagonal components. 
In the first case $i=p_+$ and $j=p_-$. We have
\begin{flalign}
(\hat{\mathbf{G}}_\Delta)_{p_+p_-}\sim (\hat{\mathbf{G}}_e)_{p_+p_+}\psi_{p_+}(\hat{\mathbf{G}}_e^0)_{p_+p_-}+(\hat{\mathbf{G}}_e)_{p_+p_-}\psi_{p_+}(\hat{\mathbf{G}}_e^0)_{p_-p_-}.
\end{flalign}
In the second case $i=p_-$ and $j=p_+$. We subsequently obtain:
\begin{flalign}
(\hat{\mathbf{G}}_\Delta)_{p_-p_+}\sim (\hat{\mathbf{G}}_e)_{p_-p_+}\psi_{p_+}(\hat{\mathbf{G}}_e^0)_{p_+p_+}+(\hat{\mathbf{G}}_e)_{p_-p_-}\psi_{p_+}(\hat{\mathbf{G}}_e^0)_{p_-p_+}.
\end{flalign}
It is important to notice that $(\mathbf{\hat{G}}_e)_{p_+p_-}$ and $(\mathbf{\hat{G}}_e)_{p_-p_+}$ transform differently. Under a $C_3$ rotation, 
\begin{flalign}
(\nonumber&\mathbf{\hat{G}}_e)_{p_+p_-}\rightarrow \omega^2(\mathbf{\hat{G}}_e)_{p_+p_-}\\
\nonumber&(\mathbf{\hat{G}}_e)_{p_-p_+}\rightarrow \omega (\mathbf{\hat{G}}_e)_{p_-p_+}\\
\nonumber &\psi_{p_+}\rightarrow \omega \psi_{p_+},
\end{flalign}
with $\omega=e^{2i\pi/3}$. As a consequence, only the product $(\hat{\mathbf{G}}_e)_{p_+p_-}\psi_{p_+}$ is invariant under the $C_3$ rotation. This has a direct consequence for the off-diagonal components of the local anomalous Green function. Specifically, $\mathbf{G}_{p_-p_+}^\Delta$ is negligible in comparison to $\mathbf{G}_{p_+p_-}^\Delta$, which indicates that the $p+ip'$ pairing is dominated by inter-orbital effects and belongs to the type 2 class. As a result, the local anomalous Green function takes the specific form in the $(p_+,p_-)$ basis
\begin{flalign}
\hat{\mathbf{G}}_{\Delta}=\begin{pmatrix}
0 & \mathbf{G}_{p_+,p_-}^\Delta \\
0 & 0 \\
\end{pmatrix}.
\end{flalign}
For a $p-ip'$ pairing, $\mathbf{G}_{p_+p_-}^\Delta$ is negligible with respect to  $\mathbf{G}_{p_-p_+}^\Delta$, and the local anomalous Green function is the transpose of the $p+ip'$ case.

Now let's treat the case of $p_x/p_y$ pairing which also belongs to the $E_1$ irreducible representation of $D_6$. Both $p_x$ and $p_y$ can be expressed as a linear combinations of $p+ip'$ and $p-ip'$ which means that the local anomalous Green function will be a combination of the two cases. It will be off-diagonal, so dominated by inter-orbital effects, but of type 1 because the local anomalous Green's functions $\hat{G}_\Delta$ have the following structure
\begin{flalign}
\hat{\mathbf{G}}_{\Delta}=\begin{pmatrix}
0 & \mathbf{G}_{p_+p_-}^\Delta \\
\mathbf{G}_{p_-p_+}^\Delta & 0 \\
\end{pmatrix}.
\end{flalign}

\vspace{1\baselineskip}

\subsubsection{$d$-wave pairing\label{sec:symmetry_dwave}}
The same analysis can be done for the $d+id'$ and $d-id'$ pairings, and then extended to $d_{xy}$ and $d_{x^2-y^2}$ pairings. The only difference is that because $d+id'$ has an angular momentum $\ell=2$, under a $C_3$ rotation,  
\begin{flalign}
\nonumber&\mathbf{G}_{p_+p_-}\rightarrow \omega ^2\mathbf{G}_{p_+p_-}\\
\nonumber&\mathbf{G}_{p_-p_+}\rightarrow \omega \mathbf{G}_{p_-p_+}\\
\nonumber &\psi_{p_+}\rightarrow \omega^2 \psi_{p_+},
\end{flalign}
with $\omega=e^{2i\pi/3}$. So the local anomalous Green's function of $d+id'$ has the same form than the one for $p+ip'$ but with the non-zero components located at $p_-,p_+$ instead of $p_+,p_-$,
\begin{flalign}
\hat{\mathbf{G}}_\Delta = \begin{pmatrix}
    0 & 0 \\
    \mathbf{G}_{p_-p_+}^\Delta & 0 \\
\end{pmatrix}.
\end{flalign}
The local anomalous Green's function for $d-id^\prime$ is the transpose of the $d+id^\prime$ one. 
%
\section{2-Band model\label{app:2bandmodel}}
The 6-band model is an effective model that is supposed to reproduce the correct low-energy band structure and the correct chirality of the Dirac cones~\cite{song2019all} originating from the two twisted layers of TBG. Other effective tight-binding models exist, in particular a 2-band model with a point group symmetry $D_3$, which has been used many times because its simplicity~\cite{de2020valley,fujimoto2021moire}. However, as noted in Sec.~\ref{sec:Normal}, the 2-band model does not capture the topology of TBG. Because of the Nielsen–Ninomiya theorem~\cite{nielsen1981no}, the two inequivalent $K$ points of graphene have opposite chirality~\cite{alvarado2021transport} contrary to TBG. Details of this model can be found in~\cite{alvarado2021transport}. 
In the following, we perform a computation of the Chern number using this simpler model. To enable a comparison with the same symmetries in the 6-band model, we considered an NNN pairing for the 2-band model, so that each site is paired with 6 other sites as in the 6-band model.
%
\subsection{Chern number}\label{app:Chern2Band}


\begin{figure}[h]
	\begin{center}
		\begin{tikzpicture}
			\node at (-5,0) {
				\includegraphics[width=9cm]{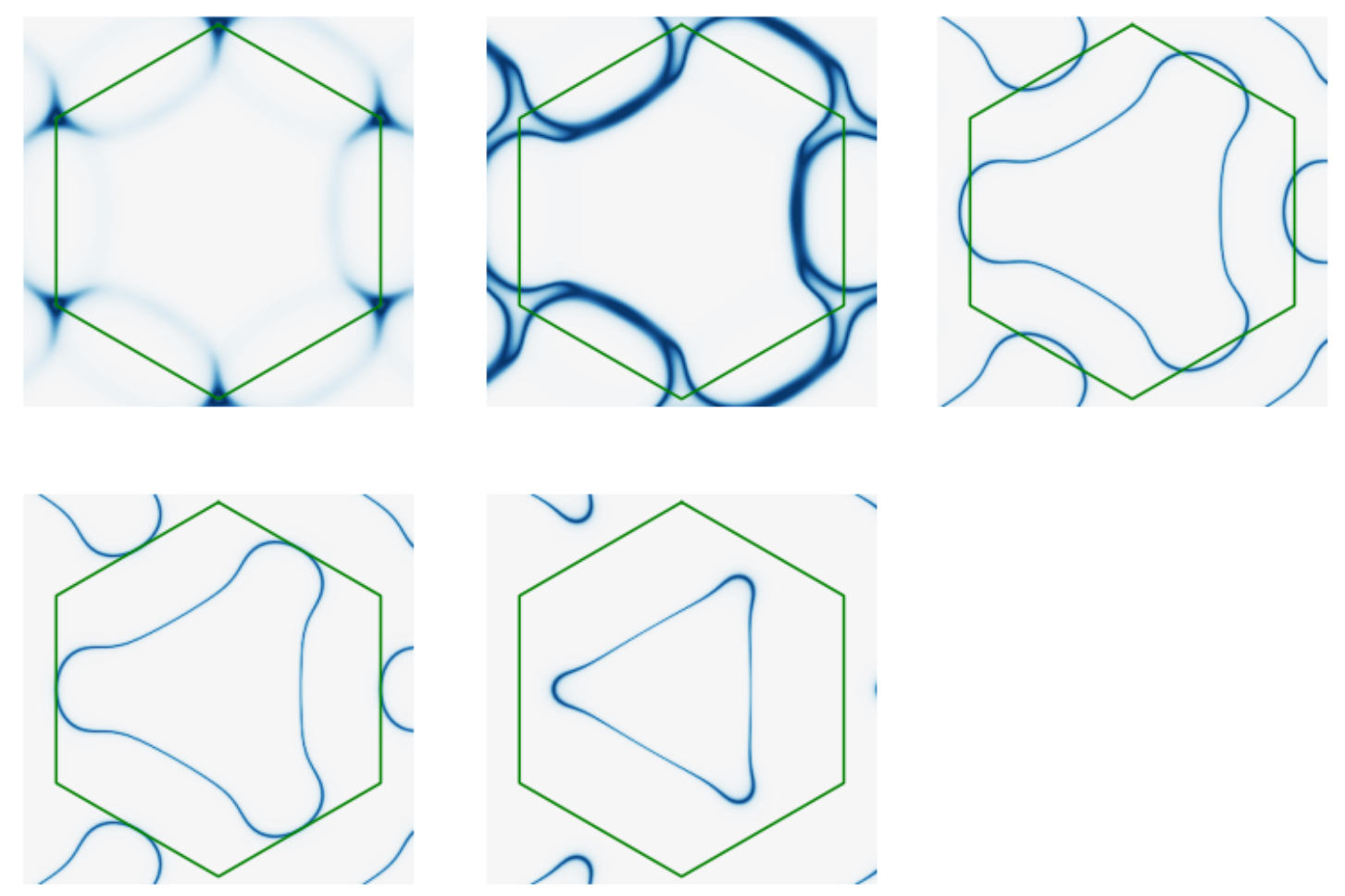}};
			\node at (-9,3.1) {(a)};
			\node at (-6,3.1) {(b)};
			\node at (-3,3.1) {(c)};
			\node at (-9,-0.1) {(d)};
			\node at (-6,-0.1) {(e)};
			%
		\end{tikzpicture}
	\end{center}
	\caption{Fermi Surface of the 2B1V model for (a) $\mu=-0.05$, (b) $\mu=-0.16$, (c) $-1.0$, (d) $\mu=-1.4$ and (e) $\mu=-2.5$. We see that at low $\mu$ the Fermi Surface is centered around the $K$ points and at larger $\mu$ it is centered around $\Gamma$. Compared with the 6-band model in the main text (Fig.~\ref{fig:FS_6B1V}), both models have the same qualitative Fermi surface evolution.}
	\label{fig:Normal_state_band2} 
\end{figure}
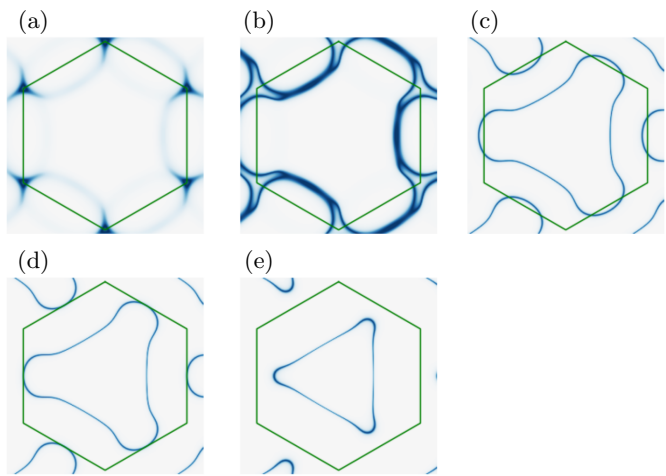
\begin{figure*}[h!]
	\begin{center}
		\begin{tikzpicture}
			\node at (0,0) {
				\includegraphics[width=14.5cm]{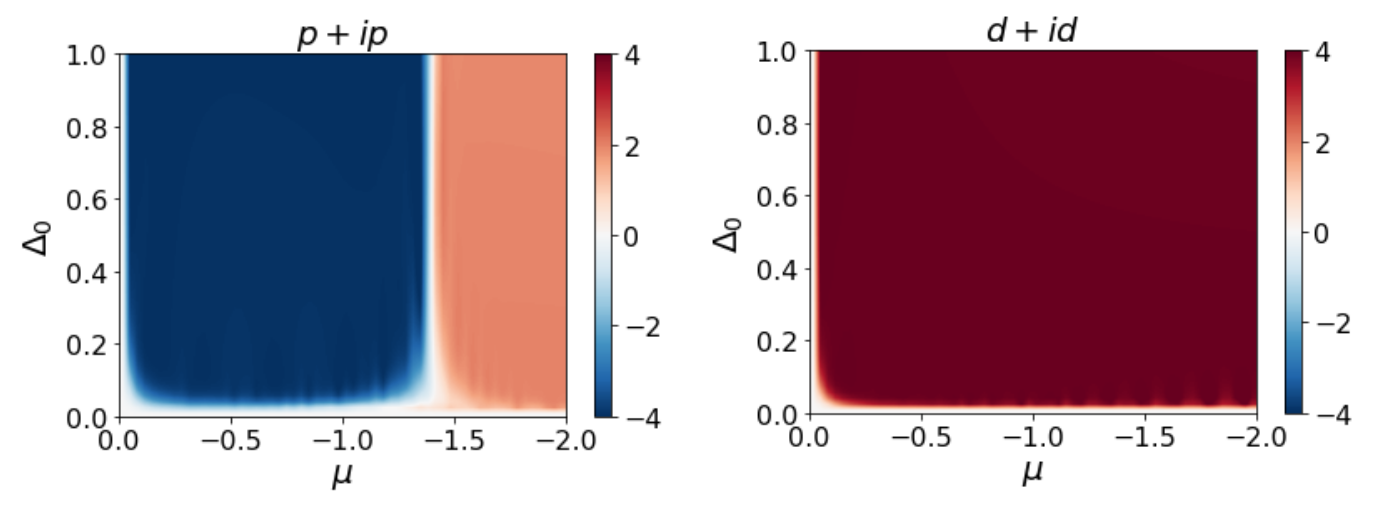}};
			\node[scale=1.3] at (-5.2,2.3) {(a)};
			\node[scale=1.3] at (2.,2.3) {(b)};
		\end{tikzpicture}
	\end{center}
	\caption{Chern number as a function of chemical potential $\mu$ and superconducting
		coupling $\Delta_0$ for the 2-Band model with (a) $p+ip'$ pairing and
		(b) $d+id'$ pairing. One has the same qualitative behavior as for the 6-band model described in the main text (Fig.~\ref{fig:Chern_6B1V}).}
	\label{fig:Chern_2B1V} 
\end{figure*}

We perform the same computation as in Sec.~\ref{sec:Chern} for the chiral pairings $p+ip'$ and $d+id'$-wave. We find that the evolution of the Chern number with respect to the chemical potential
$\mu$ and the magnitude of the superconducting order parameter, $\Delta_{0}$, is identical for the two models. Crucially, the Fermi surface of the 2-band model has qualitatively the same $\mu$ evolution as that of the 6-band model described in the main text, see Fig.~\ref{fig:Normal_state_band2}. It is therefore not surprising that the Chern number behavior is qualitatively the same for both models, see Fig.~\ref{fig:Chern_2B1V}. This results further emphasizes the significant role of the normal state topology for the SC topological phase transitions.
%
\end{document}